\documentclass[11pt]{article}

\usepackage[T1]{fontenc}
\usepackage[cp1251]{inputenc}
\usepackage{textcomp}
\usepackage[centertags]{amsmath}
\usepackage[mediummath]{nccmath}
\usepackage{amsfonts}
\usepackage{amssymb}
\usepackage{graphicx}
\usepackage{graphbox}
\usepackage[numbers,sort&compress]{natbib}
\usepackage[pdftex]{hyperref}
\usepackage{calc}

\usepackage{paperinitial}

\paperinitialization{15mm}{15mm}{15mm}{15mm}{2pt}{10pt}

\DeclareMathOperator{\im}{Im}

\DeclareMathOperator{\diag}{diag}

\newcommand{\bs}{\boldsymbol}

\newcommand{\e}{\varepsilon}
\newcommand{\vf}{\varphi}
\newcommand{\vk}{\varkappa}
\newcommand{\s}{\sigma}

\newcommand{\al}{\alpha}
\newcommand{\be}{\beta}
\newcommand{\ga}{\gamma}

\newcommand{\de}{\delta}
\newcommand{\De}{\Delta}

\newcommand{\spx}{\mathbf{x}}

\newcommand{\spk}{\mathbf{k}}
\newcommand{\spe}{\mathbf{e}}

\begin{document}
\allowdisplaybreaks[4]
\frenchspacing
\setlength{\unitlength}{1pt}

\title{{\Large\textbf{Generation of hard twisted photons by charged particles in cholesteric liquid crystals}}}
\date{}

\author{
O.V. Bogdanov${}^{1),2)}$\thanks{E-mail: \texttt{bov@tpu.ru}},\;
P.O. Kazinski${}^{1)}$\thanks{E-mail: \texttt{kpo@phys.tsu.ru}},\;
P.S. Korolev${}^{1)}$\thanks{E-mail: \texttt{kizorph.d@gmail.com}},\;
and G.Yu. Lazarenko${}^{1)}$\thanks{E-mail: \texttt{laz@phys.tsu.ru}}\\[0.5em]
{\normalsize ${}^{1)}$ Physics Faculty, Tomsk State University, Tomsk 634050, Russia}\\
{\normalsize ${}^{2)}$ Tomsk Polytechnic University, Tomsk 634050, Russia}
}

\maketitle

\begin{abstract}

We study the radiation from charged particles crossing a cholesteric plate in the shortwave approximation when the wavelength of photons is much smaller than the pitch of the cholesteric helix whereas the escaping angle of the photon and the anisotropy of the permittivity tensor can be arbitrary. The radiation of photons is treated in the framework of quantum electrodynamics with classical currents. The radiation of the plane-wave photons and the photons with definite projection of the angular momentum (the twisted photons) produced by charged particles crossing the cholesteric plate and moving rectilinearly and uniformly is considered. The explicit expressions for the average number of radiated photons and their spectra with respect to the energy and the projection of the angular momentum are obtained in this case. It turns out that in the paraxial approximation the projection of the orbital angular momentum, $l$, of radiated twisted photons is related to the harmonic number $n\in \mathbb{Z}$ as $l=2n+1$, i.e., the given system is a pure source of twisted photons as expected. It is shown that in the paraxial shortwave regime the main part of radiated photons is linearly polarized with $l=\pm1$ at the harmonics $n=\{-1,0\}$. The applicability conditions of the approach developed are discussed. As the examples, we consider the production of $6.3$ eV twisted photons from uranium nuclei and the production of X-ray twisted photons from $120$ MeV electrons.

\end{abstract}

\section{Introduction}

The vortex waves are called the electromagnetic waves with a helical phase front characterized by the phase dependence of the form $e^{il\vf}$, where $l$ is the projection of the orbital angular momentum (OAM) and $\vf$ is the azimuthal angle \cite{OAM,SerboNew,New18,New19,OAMPM,PadgOAM25,Roadmap16}. Evolving in time, this wave front twists around the average direction of propagation of the electromagnetic wave and so these waves are also referred to as twisted ones \cite{MolTerTorTor}. The amplitude of the electromagnetic field of the vortex wave vanishes on the axis of its propagation. The notion of twisted waves is generalized to the nonparaxial regime \cite{GottfYan,JaurHac,JenSerprl,JenSerepj} where the photons constituting such an electromagnetic wave possess the projection of the total angular momentum $m$ and the helicity $s$. In the paraxial regime, the twisted photons constituting a vortex wave carry the projection of OAM $l=m-s$ \cite{OAM}. Because of their peculiar properties, the twisted photons found many applications in physics, biology, and telecommunication technologies \cite{SerboNew,New18,New19,OAMPM,PadgOAM25,Roadmap16}. For example, the use of twisted photons in telecommunication increases the density of information transfer by $2^{l}$ times, the quantum number $l$ is used in quantum cryptography, whereas in optics the twisted photons were employed to overcome the diffraction limit (see the references in  \cite{SerboNew,New18,New19,OAMPM,PadgOAM25,Roadmap16}). In order to take advantage of the properties of twisted photons in various fields of fundamental science and technology, there is a pressing need in the development of new pure bright sources of twisted photons in various spectral ranges.

Nowadays there are various methods to convert a plane-wave electromagnetic radiation to a twisted one in the radio and optical spectrum ranges (see the reviews \cite{OAMPM,PM,SerboNew}). The developing new materials allow one to extend the spectrum range where such methods are applicable \cite{XRplate}. The main advantage of such methods is their simplicity. However, they possess certain drawbacks: low intensity, restrictions on the energy of created twisted photons, and poor variability of parameters of the produced radiation. These shortcomings beget the necessity in developing of alternative approaches to generation of twisted photons. The more intense sources of twisted photons do not rely on the conversion of a plane-wave radiation into twisted one but are based on the direct generation of twisted photons by charged particles. One can distinguish the main of such approaches: undulator radiation \cite{BKL2,BKL4,BKL6,BKL8,SasMcNu,BHKMSS,HKDXMHR}, the Compton effect \cite{BKL4,SerboNew,JenSerprl,JenSerepj,KatohPRL}, channeling radiation \cite{ABK,Epp}, Vavilov-Cherenkov (VC) and transition radiations \cite{Kaminer,BKL5,BKL7,BKL8,parax,IvSerZay16}. These methods enable one to create an intense hard electromagnetic radiation consisting of photons with a definite projection of the angular momentum. Larger projections of the angular momentum can be achieved by the use of the coherent radiation from helically microbunched beams of charged particles \cite{HKDXMHR,BKLb,BKL7,HemStuXiZh14}.

It was shown in the papers \cite{BKL5,BKL7} that even in the simplest case of a Gaussian beam of charged particles transition and VC radiations consist of twisted photons with nonzero projection of the OAM $l=-s$. Furthermore, employing the symmetry arguments, it was shown in the paper \cite{BKL5} that the use of helical media allows one to produce the twisted photons with higher values of OAM in the radiation from charged particles moving in such a medium. In our recent paper \cite{parax}, the theory of radiation of twisted photons developed in \cite{BKL5} was generalized to the case of anisotropic media. The approximate solutions of the Maxwell equations for cholesteric liquid crystals (CLC), which are a particular case of a helical medium, were found in that paper for the two cases: the paraxial approximation and the approximation of a small anisotropy. These solutions were used to derive the explicit expressions for the probability to detect plane-wave and twisted photons in transition radiation generated by charged particles in these regimes. In the present paper, our aim is to apply the shortwave approximation \cite{AksValRom01,AksValRom04,AksKryuRom06,Aksenova2008,BabBuld,BagTrif,RMKD18,Maslov} to describe the radiation of photons produced by charged particles moving in the CLC plate of a finite width for such parameters where the approximate approaches used in \cite{parax} are not valid.

The paper is organized as follows. In Sec. \ref{CLC}, we describe the optical properties of CLCs and discuss different approaches to solution of the corresponding Maxwell equations. Section \ref{CCM} is devoted to the shortwave approximation employed to construct the mode functions of a quantum electromagnetic field. The boundary conditions on the interfaces of the CLC plate are taken into account. In Sec. \ref{PPP}, we obtain the explicit expressions for the average number of plane-wave photons radiated by a charged particle moving in the CLC plate and investigate the spectral properties of this radiation. In Sec. \ref{ProbTwPhWKB}, we deduce the average number of radiated twisted photons using the expansion of the plane-wave photons in terms of the twisted ones. The selection rules resulting from a peculiar structure of CLCs readily follow from the explicit expression for the radiation probability of twisted photons. For the reader convenience some lengthy formulas are removed to Appendices \ref{A}, \ref{Join_Coeff_App}, \ref{C}.

We use the system of units such that $\hbar=c=1$ and $e^2=4\pi\al$, where $\al$ is the fine structure constant. We also suppose $(x,y,z)\equiv(x_1,x_2,x_3)$ everywhere in the text.

\section{Electromagnetic properties of cholesterics}\label{CLC}

At present, the liquid crystals are well studied and find various applications in science and technology \cite{Andri,Vetrov,Chandra77,deGennProst,BelBook}. For our purpose, the CLCs are of a particular interest since the director forms a helical structure in these nematic liquid crystals \cite{Andri,BelBook,Vetrov}. This feature allows one to use the CLCs for generation of twisted photons, which are radiated from charged particles passing through such a helical structure \cite{BKL5,parax}. The other means of conversion of plane-wave photons into twisted ones with the aid of liquid crystals including CLCs were proposed in \cite{CLCTP,Barboza1,Barboza2}. The peculiar electromagnetic properties of CLCs are described by the permittivity tensor of the form \cite{BelBook,LL8,Vetrov,Chandra77,deGennProst}
\begin{equation}\label{permit_holec}
    \e_{ij}(k_0,z)=\e_\perp(k_0)\de_{ij}+(\e_\parallel(k_0)-\e_\perp(k_0))\tau_{i}(z)\tau_{j}(z), \qquad \bs{\tau}(z)=(\cos(qz),\sin(qz),0),
\end{equation}
where $\e_{\parallel}$ is the permittivity along the director $\bs{\tau}(z)$ and $\e_{\perp}$ is the permittivity in the direction perpendicular to it. The period of variations of the director, $2\pi/q$, is the helix pitch of the CLC. For the different CLCs, it varies from dozens of angstroms to several micrometers \cite{BelBook}. The period of variations of the electromagnetic properties of a CLC equals a half of the helix pitch \eqref{permit_holec}. The applied electric field, changes of the temperature and the pressure vary the parameters of CLCs that makes their electromagnetic properties highly adjustable \cite{Andri,BelBook,Vetrov,Chandra77,deGennProst}.

\begin{figure}[t]
   \centering
   \includegraphics*[width=0.7\linewidth]{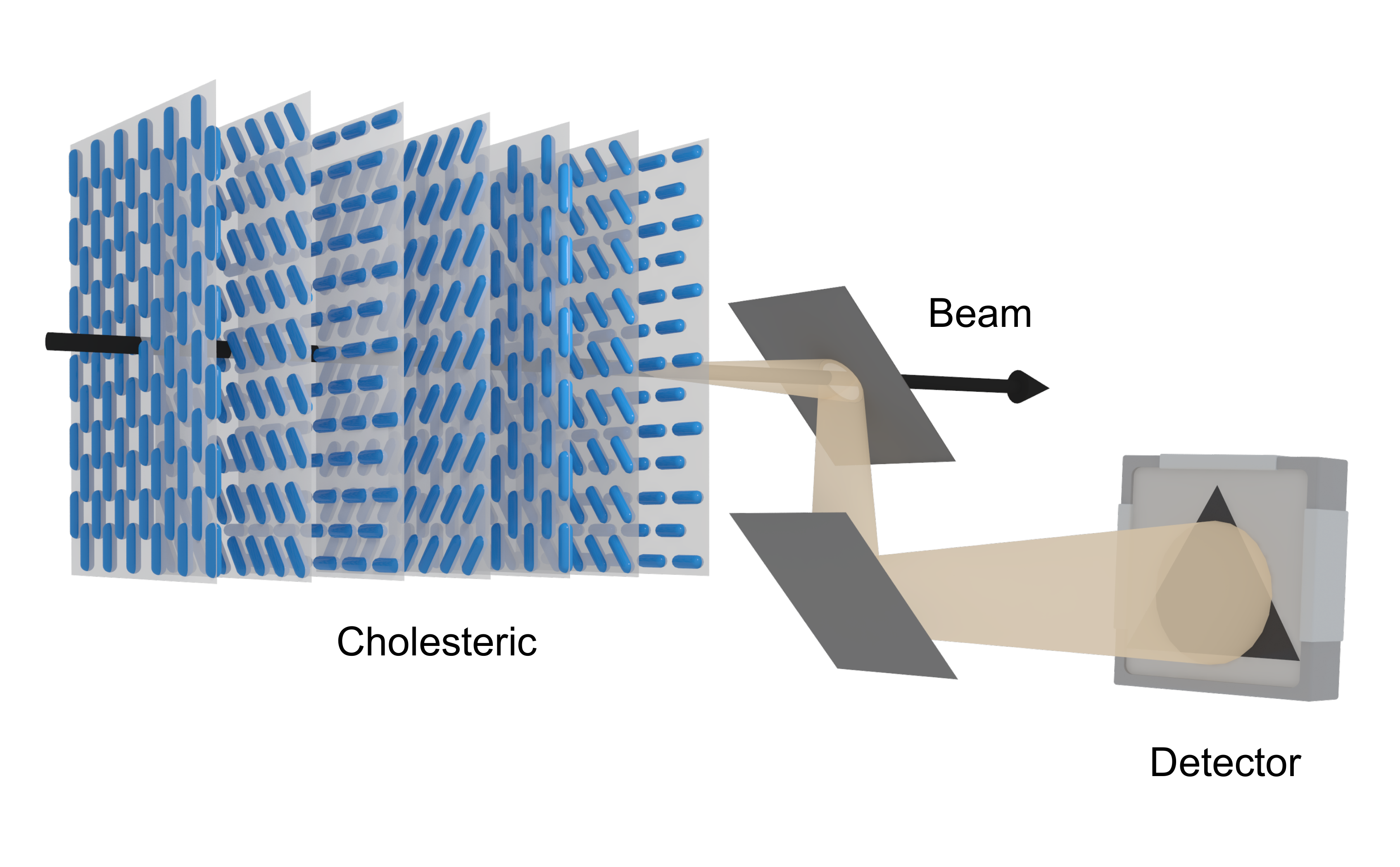}
   \caption{{\footnotesize The scheme of the experimental setup. The beam of charged particles falls normally onto the CLC plate and then passes through the perforated mirror. The transition radiation created in the CLC plate is reflected by this mirror and is brought out to the detector. The second mirror is placed parallel to the first one so that the spectrum of radiation over $m$ is preserved. Notice that the CLC is depicted only schematically. Actually, CLCs possess neither sublayers nor periodic stricture in these sublayers.}}
\label{scheme_plots}
\end{figure}

The Maxwell equation for the electromagnetic potential in a CLC are written as
\begin{equation}\label{Max_eqns}
    (k_0^2\e_{ij}(k_0,z)-\partial_i \partial_j+\De\de_{ij})A_j(\spx)=0,\qquad\partial_i(\e_{ij}(k_0,z)A_j(\spx))=0,
\end{equation}
where $k_0$ is the energy of photons in the electromagnetic wave. The second equation in \eqref{Max_eqns} is a consequence of the first one at $k_{0}\neq0$. It is a generalization of the Coulomb gauge for anisotropic media. We assume further that the boundary conditions for Eqs. \eqref{Max_eqns} are such that Eqs. \eqref{Max_eqns} do not have solutions for $k_0=0$.

The permittivity tensor \eqref{permit_holec} is invariant under translations in the plane $(x,y)$. Therefore, the solutions to \eqref{Max_eqns} can be sought in the form
\begin{equation}\label{pln_waves}
    \mathbf{A}(\spx)=e^{i\spk_\perp\spx_\perp}\mathbf{A}(z),
\end{equation}
where $\spk_\perp:=(k_1,k_2)$ and $\spx_\perp=(x_1,x_2)$. Furthermore, the tensor \eqref{permit_holec} possesses the helical symmetry, i.e., it is invariant with respect to translations along the $z$ axis and simultaneous rotations around this axis to the corresponding angle. In consequence of this symmetry, it is useful to pass from the Cartesian basis $\{\spe_1,\spe_2,\spe_3\}$ to the basis $\{\spe_+,\spe_-,\spe_3\}$, where $\spe_\pm:=\spe_1\pm i \spe_2$. Any vector $\mathbf{A}$ is written as
\begin{equation}\label{cylin_basis}
    \mathbf{A}=\frac{1}{2}(\spe_+ A_-+\spe_- A_+)+\spe_3 A_3, \qquad A_3=\spe_3\mathbf{A},\quad A_\pm=\spe_\pm\mathbf{A},
\end{equation}
in this basis. Making use of Eqs. \eqref{Max_eqns}, the third component of $\mathbf{A}$ is expressed through the rest two ones
\begin{equation}\label{a_3_comp}
    A_3=\frac{ik_\perp}{2\bar{k}_3^2}\partial_3(a_+ +a_{-}),\qquad \bar{k}_3^2:=\bar{k}_0^2-k_\perp^2,
\end{equation}
where $\bar{k}_0^2:=\e_\perp k_0^2$, $k_\perp:=|k_+|$, and
\begin{equation}
    a_\pm=A_\pm e^{\mp i\vf},\qquad \vf:=\arg k_+.
\end{equation}
In order to find $a_\pm$, it is necessary to solve the matrix Schr\"{o}dinger equation
\begin{equation}\label{Max_eqns2}
    (\partial_3K\partial_3+V)
        \left[
      \begin{array}{c}
        a_+ \\
        a_- \\
      \end{array}
    \right]=0.
\end{equation}
Here
\begin{equation}
    K=\left[
      \begin{array}{cc}
         1+\frac{k_\perp^2}{2\bar{k}_3^2} & \frac{k_\perp^2}{2\bar{k}_3^2} \\
         \frac{k_\perp^2}{2\bar{k}_3^2} & 1+\frac{k_\perp^2}{2\bar{k}_3^2} \\
      \end{array}
    \right],\qquad
    V=-
    \frac{k_\perp^2}{2}
    \left[
      \begin{array}{cc}
         1 & -1 \\
         -1 & 1 \\
      \end{array}
    \right]+
    \frac{\bar{k}_0^2}{2}
    \left[
      \begin{array}{cc}
        2+\de\e & \de\e e^{2i\bar{\theta}} \\
        \de\e e^{-2i\bar{\theta}} & 2+\de\e \\
      \end{array}
    \right],
\end{equation}
where $\bar{\theta}:=qz-\vf$ and $\de\e:=\e_{\parallel}/\e_\perp-1$. The quantity $\bar{\theta}$ equals the angle between the director $\bs{\tau}$ and the component of the wave vector $\mathbf{k}_\perp$. The quantity $\de\e$ characterizes the anisotropy of dielectric permittivity of the CLC. It can take positive or negative values and, as a rule, is small. In the optical range, its modulus is of order $10^{-2} - 10^{-1}$ \cite{BelBook}.

\begin{figure}[!ht]
   \centering
\begin{tabular}{cc}
\includegraphics*[width=0.47\linewidth]{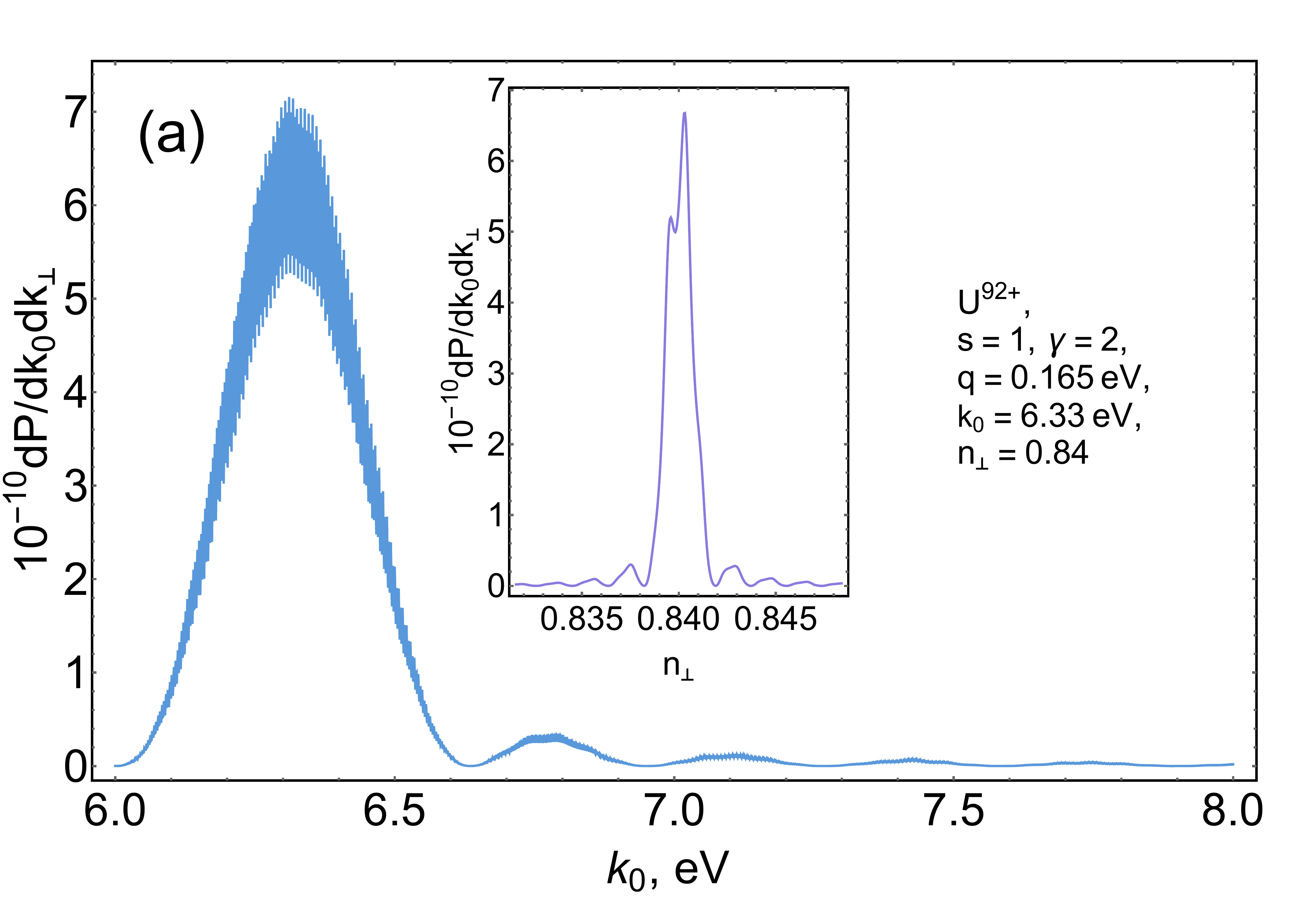}&
\includegraphics*[width=0.47\linewidth]{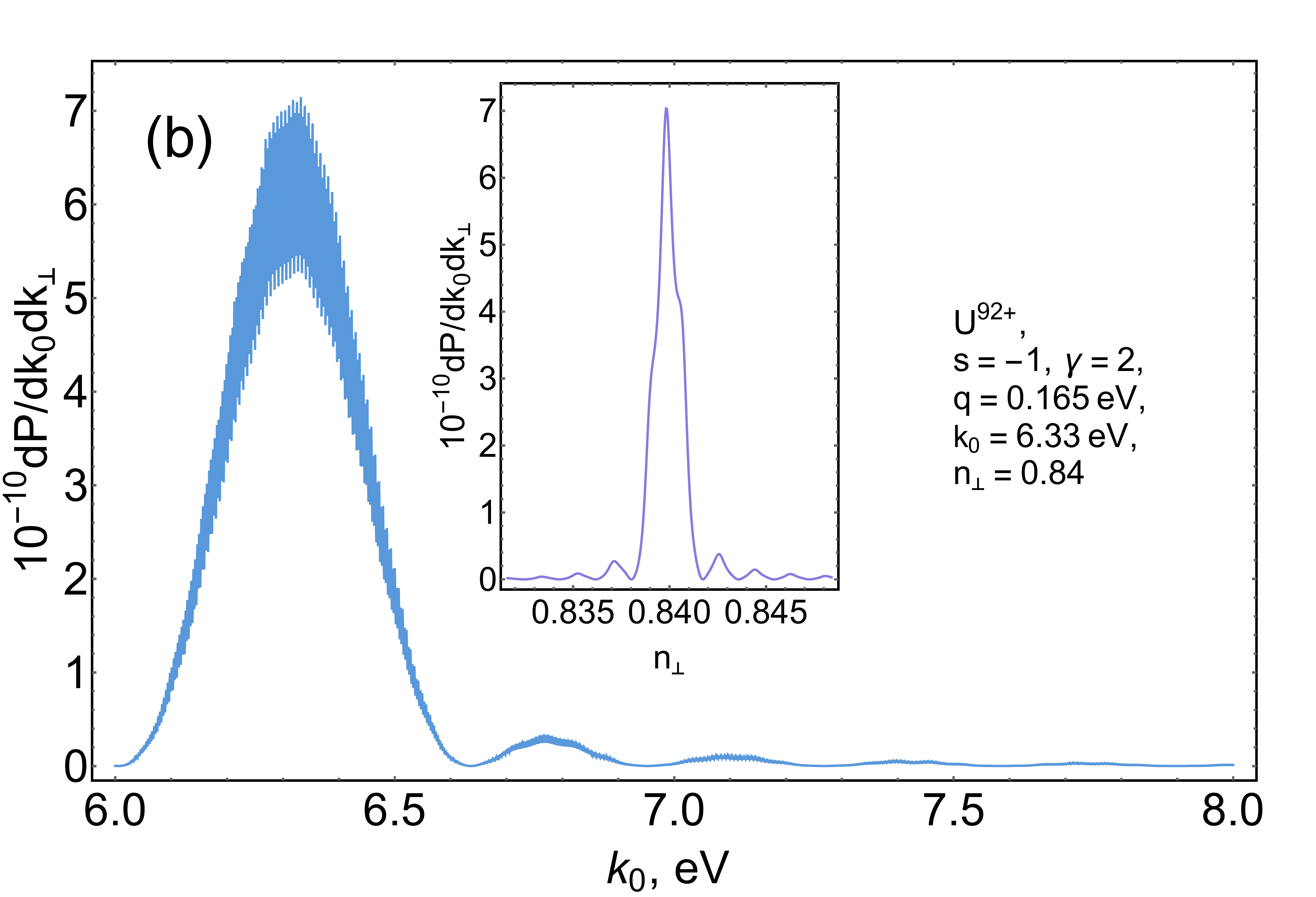}\\
\includegraphics*[width=0.47\linewidth]{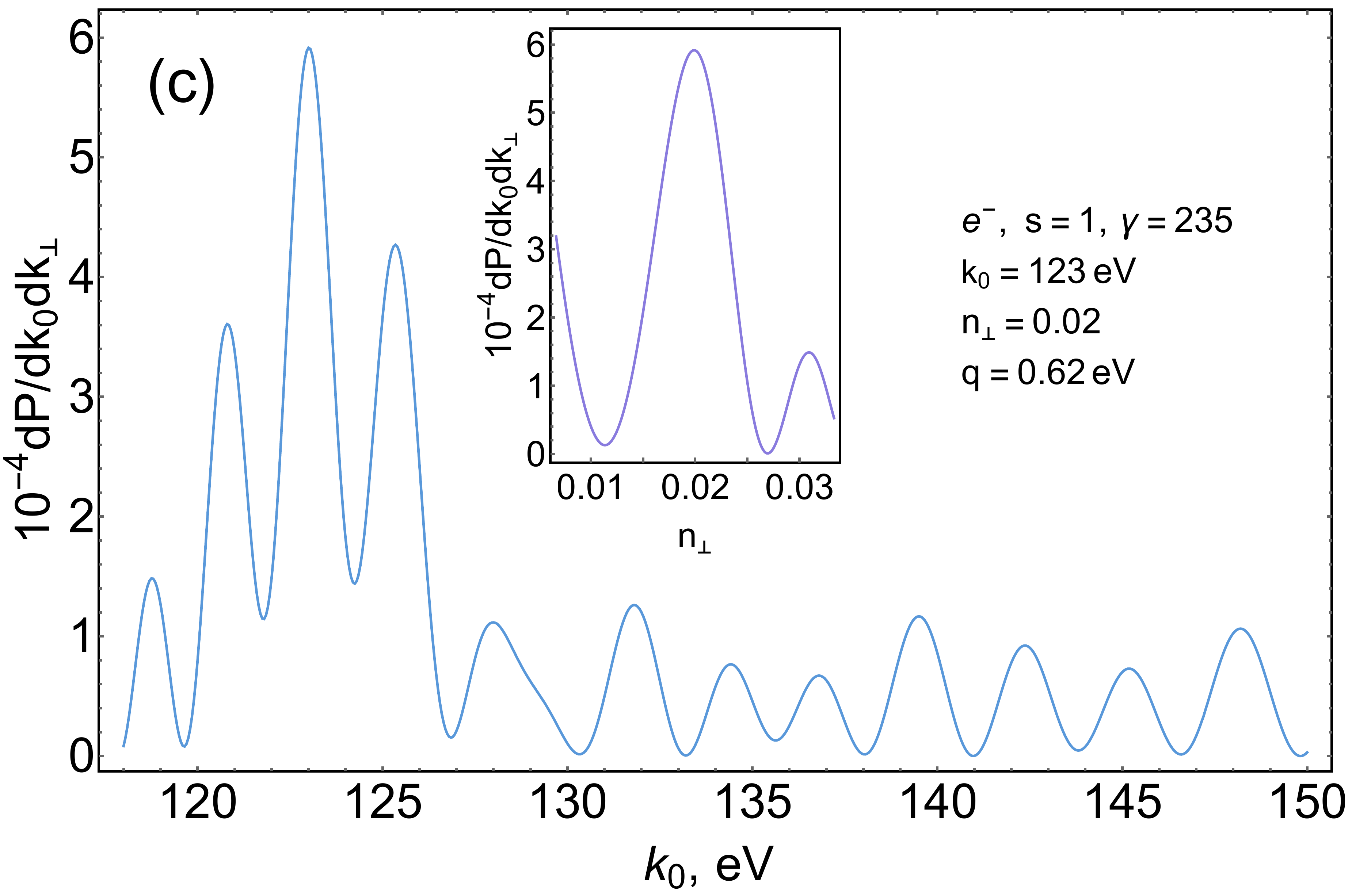}&
\includegraphics*[width=0.47 \linewidth]{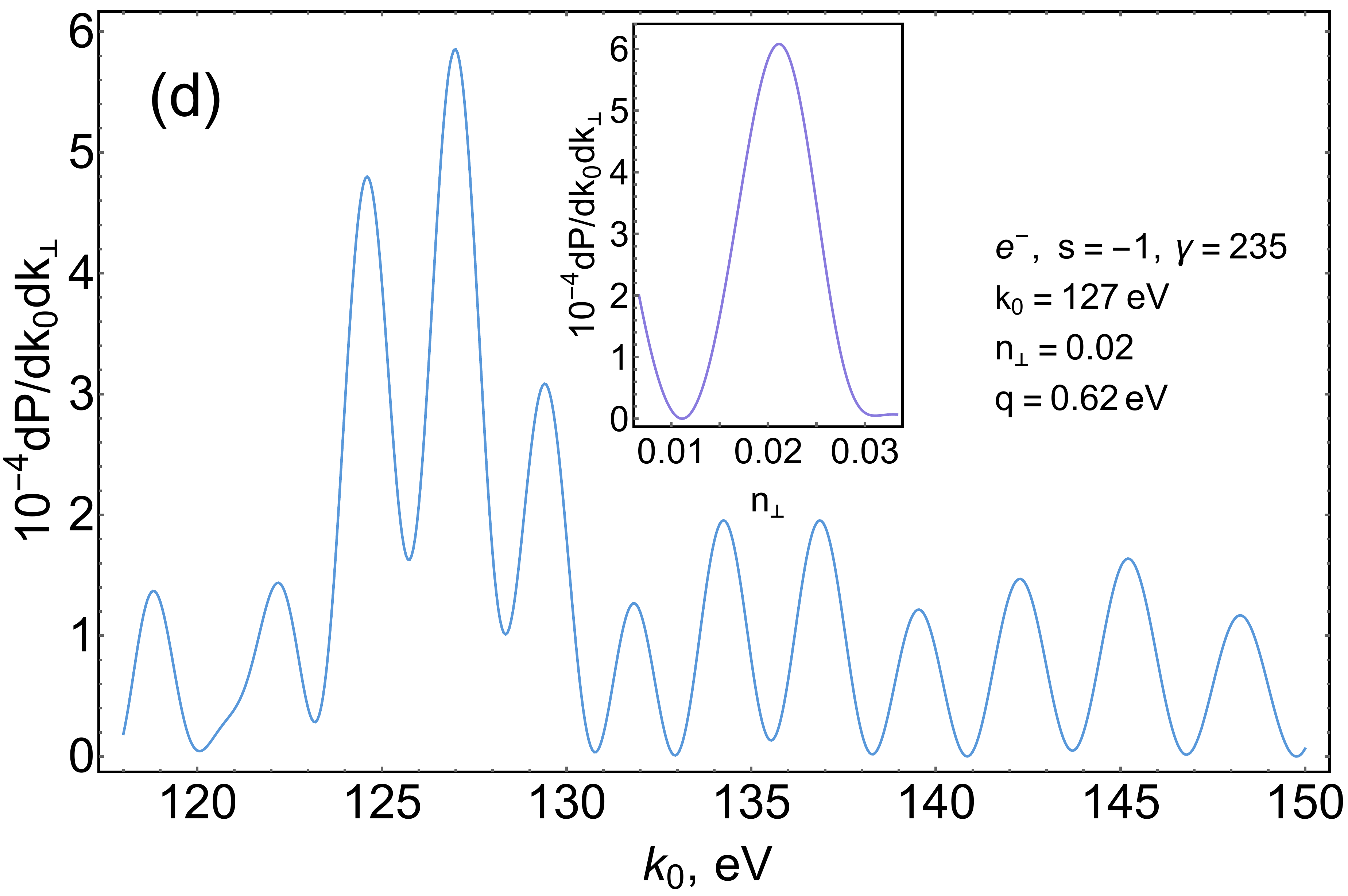}\\
\end{tabular}
    \caption{{\footnotesize The average number of plane-wave photons, $dP/dk_0dk_\perp$, produced in transition radiation by charged particles traversing normally the cholesteric plate. The shortwave approximation is used. The helicity of radiated photons is denoted as $s$ and $n_\perp=k_\perp/k_0$. The waviness of the plots reflects the presence of resonances in the photon spectrum of the CLC plate. The insets on the plots represent the distribution over $n_\perp$ at the maximum of radiation intensity. Upper plots: Transition radiation from the ions ${}^{238}$U${}^{92+}$ with the Lorentz factor $\ga=2$ \cite{GSI}. The plot (a) is for $s=1$ and the plot (b) is for $s=-1$. The width of the CLC plate $L=150$ $\mu$m, the number of periods $N_u=40$, and the components of the permittivity tensor are $\e_\perp=2.1$ and $\e_\parallel=2.49$ (see, e.g., \cite{LWGLW}). The parameter of applicability of the shortwave approximation standing on the left-hand side of \eqref{short_wave2} is $6.6$ at the maximum of radiation intensity. Notice that the perturbation theories with respect to $k_\perp$ and $\de\e$ are not valid for these parameters. The maxima of the corresponding applicability parameters -- the expressions on the left-hand sides of Eqs. \eqref{applic_conds_parax} and \eqref{appl_cond_de} -- are $3.5$ and $4.2$, respectively. Lower plots: Transition radiation from electrons with the Lorentz factor $\ga=235$. The plot (c) is for $s=1$ and the plot (d) is for $s=-1$. The width of the CLC plate $L=40$ $\mu$m, the number of periods $N_u=40$, and the components of the permittivity tensor are $\e_\perp=1-\omega_p^2/(3 k_0^2)$ and $\e_\parallel=1-\omega_p^2/k_0^2$ with $\omega_p=21$ eV (see for details \cite{band}). The parameter \eqref{short_wave2} is $0.89$ at the maximum of radiation intensity. Therefore, the shortwave approximation describes radiation only qualitatively. The perturbation theory with respect to $\de\e$ is not valid in this case since the maximum of the applicability parameters \eqref{appl_cond_de} is $2.3\times 10^3$. The perturbation theory with respect to $n_\perp$ works well in this parameter domain. The maximum of the applicability parameters \eqref{applic_conds_parax} is $1.5\times 10^{-3}$.}}
\label{eU_wkb_plane_plots}
\end{figure}

It seems impossible to find a general solution of the system of equations \eqref{Max_eqns2} in a closed form for arbitrary values of the parameters. The solution of \eqref{Max_eqns2} describing the electromagnetic wave propagating along the CLC axis is given in \cite{Mauguin11,deVries51,Kats}. This solution cannot be immediately used for description of radiation of twisted photons since their amplitudes vanish at $k_{\bot}=0$. The other limiting case where the solutions to \eqref{Max_eqns2} can be found is the isotropic medium, i.e., $\de\e=0$. The perturbative solutions of the system \eqref{Max_eqns2} with respect to $\de\e$ and $n_{\bot}:=k_\perp/k_0$ were considered in many papers \cite{parax,Aksenova2008,BelBook,Orlov,Shipov1,Shipov2,BelVC,Shipov3,CarlosVC,CarlosTR}. We will employ the approach based on the shortwave approximation \cite{AksValRom04,BabBuld,BagTrif,RMKD18,Maslov} allowing to find the solution to \eqref{Max_eqns2} in the regions of the parameter space where the perturbation theories with respect to $n_{\bot}$ or $\de\e$ are not applicable. To wit, it was shown in \cite{parax} that the perturbation theory with respect to $n_\perp$ is valid when
\begin{equation}\label{applic_conds_parax}
\begin{aligned}
    \frac{k_\perp^2}{\bar{k}_0^2}&\ll1,&\qquad\frac{k_\perp^2}{\bar{k}_0^2}\frac{|q|}{|\bar{k}_0-q|}&\ll1, \qquad\frac{k_\perp^2}{|\bar{k}_0-q||\bar{k}_0-2q|}\frac{\de\e^2\bar{k}_0^2}{256 q^2}\ll1,&\quad&\text{for $\frac{\de\e^2\bar{k}_0^2}{16 q^2}\ll1$};\\
    \frac{k_\perp^2}{\bar{k}_0^2}&\ll1,&\qquad\frac{|\de\e|\bar{k}_0}{16|q|}\frac{k_\perp^2}{\bar{k}_0^2}&\ll1,& \quad&\text{for $\frac{\de\e^2\bar{k}_0^2}{16 q^2}\gg1$};
\end{aligned}
\end{equation}
where it is assumed that $|\de\e|\ll1$. As for the perturbation theory with respect to $\de\e$, $|\de\e|\ll1$, it was found in \cite{parax} that it is applicable in the parameter domain
\begin{equation}\label{appl_cond_de}
    \Big|\frac{\de\e}{4}\frac{\e_\perp}{\e_\perp-n_\perp^2}\Big|\ll1,\qquad \Big|\frac{\de\e\bar{k}^5_3}{4qk_\perp^2 (\bar{k}_3^2-q^2)}\Big|\ll1.
\end{equation}
In particular, for $\bar{k}_3^2\gg q^2$ the last condition in \eqref{appl_cond_de} implies
\begin{equation}\label{appl_cond_he_de}
    \Big|\frac{\de\e\bar{k}^3_3}{4qk_\perp^2}\Big|\ll1.
\end{equation}
The left-hand side of this inequality grows in increasing the photon energy.

\section{Shortwave approximation}\label{CCM}

In order to obtain the approximate solutions to Eqs. \eqref{Max_eqns2}, we shall use the standard procedure of the shortwave approximation for matrix equations (see, e.g., \cite{BabBuld,BagTrif,Aksenova2008,RMKD18,Maslov}). In the papers \cite{AksValRom01,AksValRom04,AksKryuRom06,Aksenova2008}, this method was applied to description of propagation of the electromagnetic waves in CLCs. Unfortunately, in those papers, these solutions are not presented in the form that is needed for construction of a quantum electromagnetic field in the CLC plate of a finite width. In particular, the explicit form of the coefficients of linear combination of solutions to \eqref{Max_eqns2} that provide a joining with the solutions of the Maxwell equations in a vacuum is not given in those papers for arbitrary parameters of the electromagnetic wave. Therefore, for the reader convenience and conformity of notation, we derive these solutions in this section and construct the quantum electromagnetic field in the CLC plate of a finite width employing the procedure given in \cite{BKL5,parax}.

We seek for the solution of \eqref{Max_eqns2} in the form of the asymptotic series in powers of $1/k_0$:
\begin{equation}\label{WKB_expans}
    \Psi(z):=
    \left[
      \begin{array}{c}
        a_+ \\
        a_- \\
      \end{array}
    \right]
    =M(z)e^{iS(z)}\simeq \sum_{n=0}^{\infty} k_0^{-n}\Psi_{n}(z)e^{iS(z)},
\end{equation}
where $S(z)$ is a scalar function of the first degree of homogeneity with respect to $k_0$. Notice that the matrix $K$ is of zero homogeneity degree with respect to $k_0$ and $V$ is of second degree. Substituting the expansion \eqref{WKB_expans} into \eqref{Max_eqns2} and gathering the terms of the same homogeneity degree with respect to $k_0$, we arrive at the infinite chain of equations. The first two equations are
\begin{equation}\label{WKB_eqns1}
    \big[-(S')^2K+V\big]\Psi_{0}=0,\qquad \big[-(S')^2K+V\big]\Psi_{1} +i k_{0}K \big(2 S' (\Psi_{0})' +S'' \Psi_{0}\big)=0.
\end{equation}
In fact, the first equation is the equation for the eigenvectors corresponding to the zero eigenvalue of the matrix standing in the square brackets. Requiring that this equation possesses a nontrivial solution $\Psi_{0}(z)$, we find the semiclassical momenta $S'(z)$. Then taking into account that the matrix in the square brackets in the second equation in \eqref{WKB_eqns1} is Hermitian, we multiply this equation by $\Psi_{0}^\dag$ from the left. Separating the real and imaginary parts in the equation obtained, we arrive at
\begin{equation}\label{WKB_eqns2}
    (S' \Psi_{0}^\dag K \Psi_{0})'=0,\qquad \im(\Psi_{0}^\dag K \Psi'_{0})=0.
\end{equation}
As follows from the Fredholm theorem (see, e.g., \cite{RSv1}, Sec. VI.5), these equations are the necessary and sufficient conditions for solubility of the second equation in \eqref{WKB_eqns1}. The first equation in \eqref{WKB_eqns1} determines $\Psi_{0}(z)$ up to multiplication by an arbitrary function of $z$. The equations \eqref{WKB_eqns2} remove this ambiguity. As a result, $\Psi_{0}(z)$ are defined up to multiplication by a constant.

Consider the other eigenvectors of the matrix appearing in the first equation in \eqref{WKB_eqns1}:
\begin{equation}
    \big[-(S')^2K+V\big] \Psi_\perp=\nu \Psi_\perp,\qquad \Psi_\perp^{\dag} \Psi_{0}=0,
\end{equation}
where $\nu\neq0$. Then it follows from the second equation in \eqref{WKB_eqns1} that
\begin{equation}
    \Psi_{1}=-\frac{i k_{0}}{\nu}\frac{\big[\Psi_\perp^\dag K (2 S' \Psi'_{0} +S'' \Psi_{0})\big]}{\Psi_\perp^\dag \Psi_\perp} \Psi_\perp,
\end{equation}
where we have taken into account that the system \eqref{WKB_eqns1} is two dimensional and suppose that
\begin{equation}
    \Psi_{0}^\dag \Psi_{1}=0.
\end{equation}
In what follows, we will consider only the leading contribution to the expansion \eqref{WKB_expans}. This approximation proved to be adequate for the study of optical properties of CLCs with large pitch and is in a good agreement with the experiments \cite{AksKryuRom06,Aksenova2008,Aksenova05}.

The shortwave approximation holds if
\begin{equation}\label{short_wave1}
    k_0/|q|\gg 1.
\end{equation}
Moreover, as long as we keep the terms proportional to $\de\e$ entering into the potential $V$ as the leading contributions, the more stringent condition should be satisfied
\begin{equation}\label{short_wave2}
    k_0 |\de\e|/|q|\gg 1.
\end{equation}
Therefore, the semiclassical mode functions derived below are not applicable in the limit $\de\e\rightarrow0$. On the other hand, as is seen from \eqref{appl_cond_he_de}, the standard perturbation theory with respect to $\de\e$ is not valid in the region \eqref{short_wave2}. As far as the perturbation theory with respect to $n_\perp^2$ is concerned, it follows from the second line of \eqref{applic_conds_parax} that it does not hold in the domain \eqref{short_wave2} provided
\begin{equation}
    k_0 |\de\e|  n_\perp^2 \gtrsim 16|q|\e^{1/2}_\perp.
\end{equation}
For high photon energies, the components of the permittivity tensor behave as
\begin{equation}
    \e_{\perp,\parallel}(k_0)=1-\omega_{\perp,\parallel}^2/k_0^2,
\end{equation}
where $\omega_{\perp,\parallel}$ are the plasma frequencies. Hence, the condition \eqref{short_wave2} is violated for very high energies of photons. In that region of the parameter space, the perturbation theories with respect to $\de\e$ and $n_\perp^2$ work well, or one can apply the shortwave approximation regarding the terms proportional to $\de\e$ in \eqref{Max_eqns2} as small corrections.

The solutions of the form \eqref{WKB_expans} have a clear physical interpretation. The vector $M(z)$ found perturbatively by means of the above procedure determines the polarization of the wave and possible small phase corrections to its components. The Hamilton-Jacobi action (the eikonal $S(z)$) is responsible for the common rapidly varying phase of the wave. The general solution provided by this approach is a linear combination of the waves with different, in general, vectors $M(z)$ and Hamilton-Jacobi actions $S(z)$.

\begin{figure}[t]
   \centering
\begin{tabular}{cc}
\includegraphics*[width=0.47\linewidth]{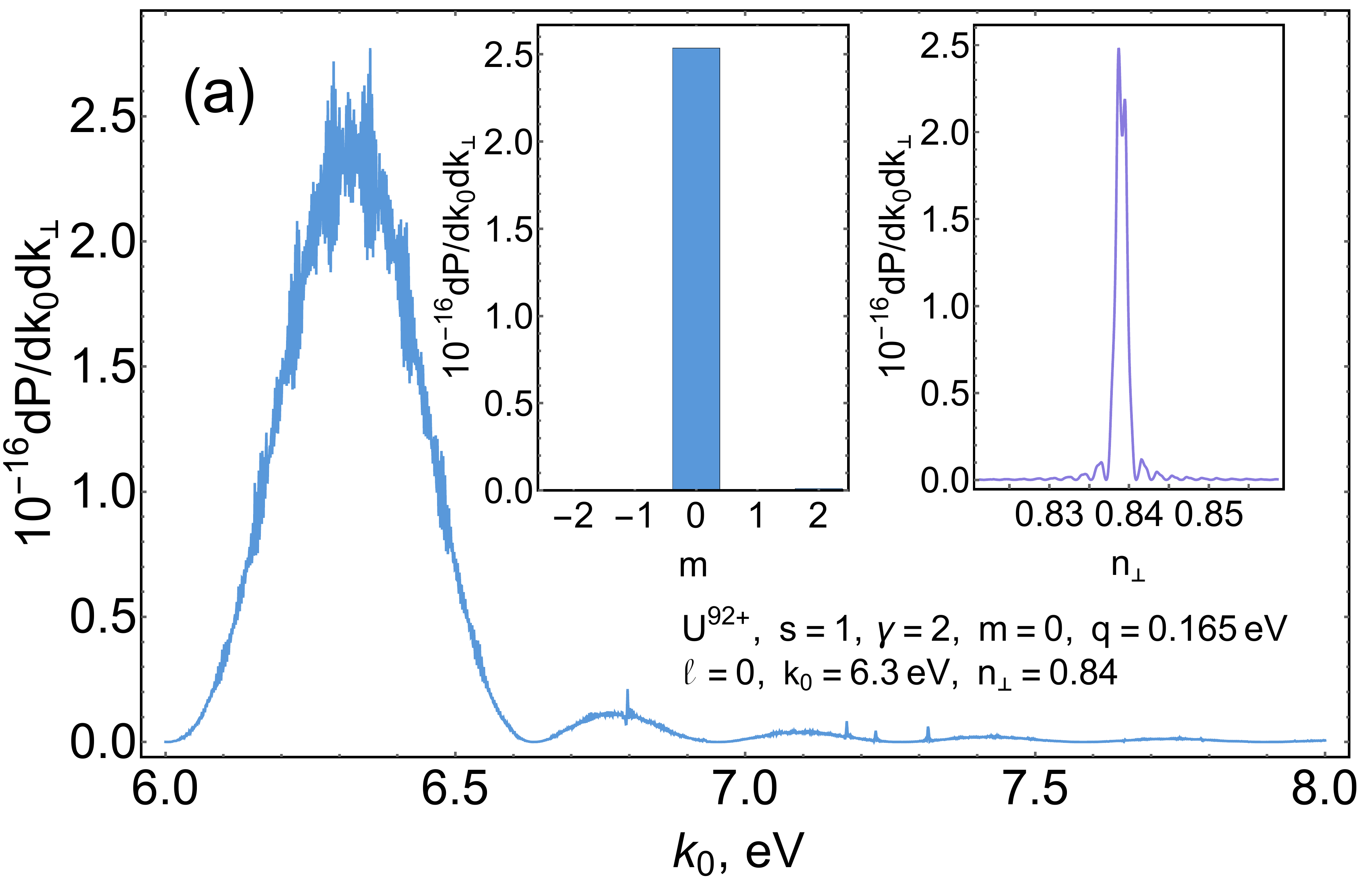}&
\includegraphics*[width=0.47 \linewidth]{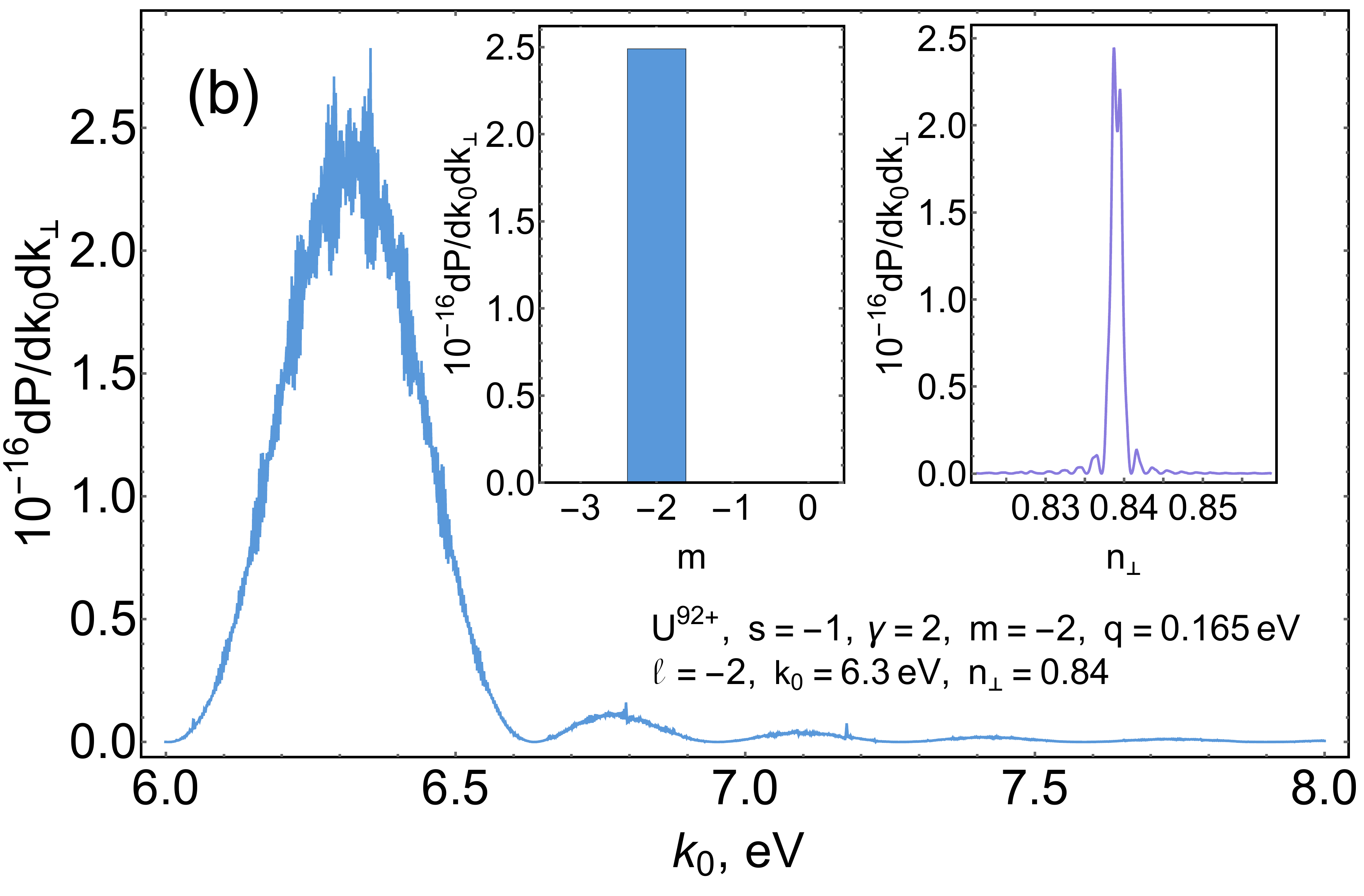}\\
\includegraphics*[width=0.47\linewidth]{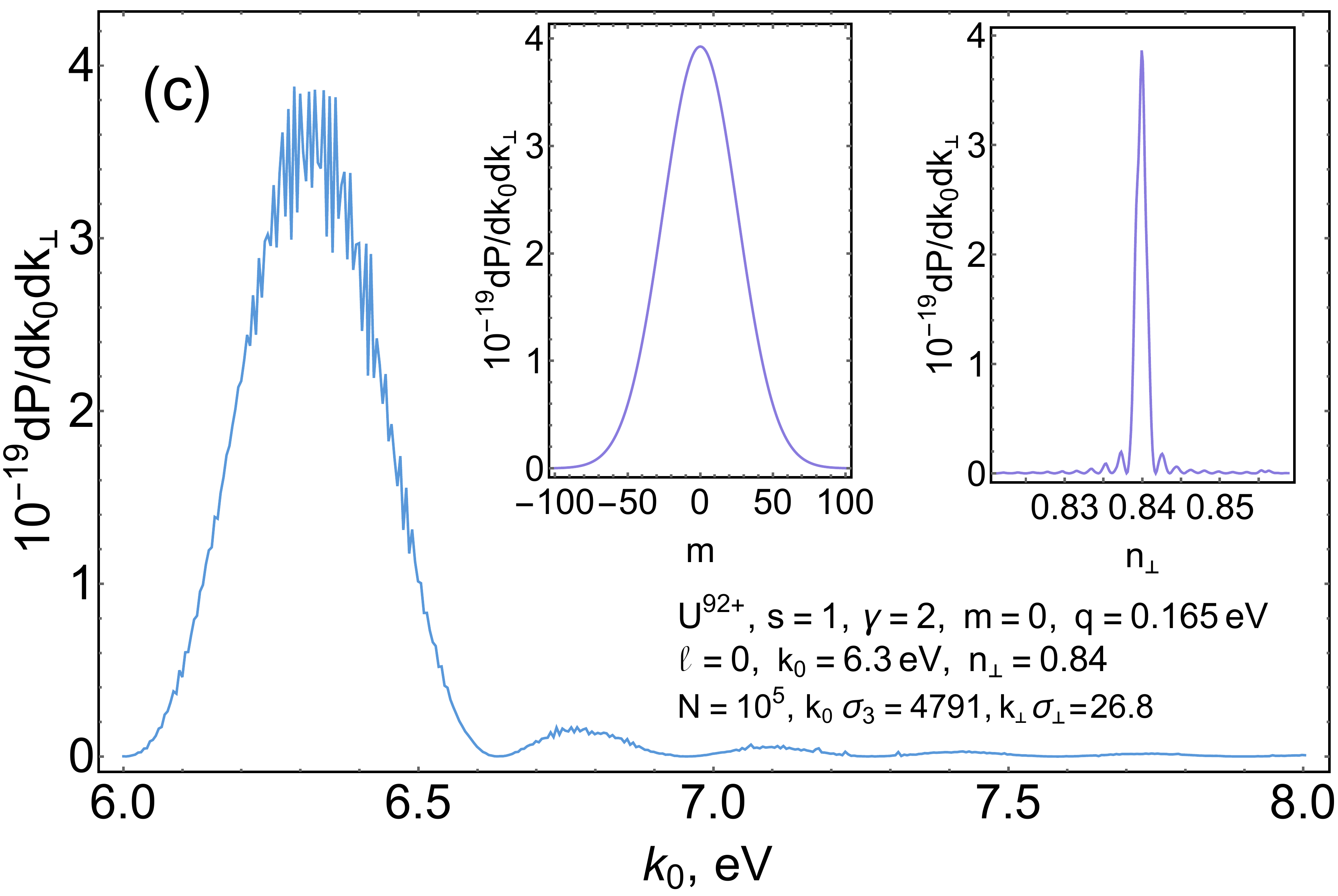}&
\includegraphics*[width=0.47\linewidth]{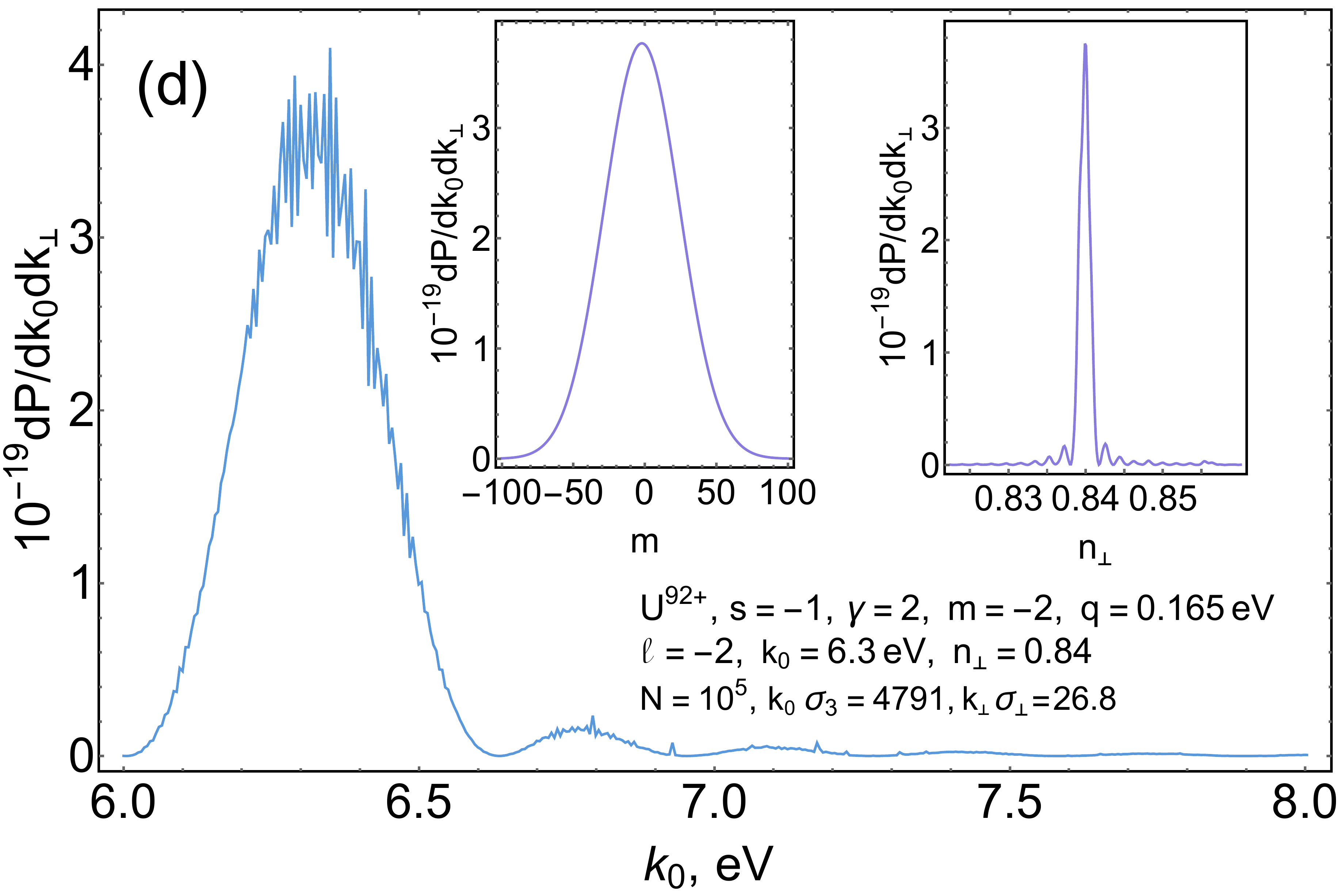}\\
\end{tabular}
    \caption{{\footnotesize The average number of twisted photons, $dP/dk_0dk_\perp$, produced in transition radiation from the ions ${}^{238}$U${}^{92+}$ traversing normally the cholesteric plate. The shortwave approximation is used. The projection of the total angular momentum is denoted as $m$. The insets on the plots represent the distributions over $m$ and $n_\perp$ at the maximum of radiation intensity. The parameters are the same as in Fig. \ref{eU_wkb_plane_plots}. The contributions with $s=1$ and $s=-1$ are almost the same indicating that the radiation polarization is linear. Upper plots: Transition radiation from the single ion ${}^{238}$U${}^{92+}$. The plot (a) is for $s=1$ and the plot (b) is for $s=-1$.  The selection rule \eqref{selection_rule} is fulfilled for $n=-1$. The negative radiation harmonic is realized because $n_\perp$ is inside of the Cherenkov cone \eqref{VCh_cone}. Lower plots: Transition radiation from the Gaussian beam of $N=10^5$ ions with transverse and longitudinal dimensions $\s_\perp=1$ $\mu$m and $\s_3=125$ $\mu$m, respectively. The plot (c) is for $s=1$ and the plot (d) is for $s=-1$. As expected \cite{BKb,BKLb}, the distribution over $m$ becomes much wider than for the radiation from a single ion since $k_\perp\s_\perp\gg1$. Nevertheless, the projection of the total angular momentum per photon, $\ell$, is the same as for the one-particle radiation.}}
\label{U_wkb_plots}
\end{figure}

Applying the above general formulas to Eq. \eqref{Max_eqns2}, we obtain the four linear independent solutions: the direct ordinary wave
\begin{equation}\label{ordinary_+}
    a^{(1)}_{\pm}=\frac{\pm i\bar{k}_{3} e^{i\bar{k}_{3}z\pm i\bar{\theta}}}{\sqrt{2\bar{k}_3(\bar{k}_0^2-k_\perp^2\cos^{2}\bar{\theta})}},\qquad A_3^{(1)}=\frac{k_\perp \sin\bar{\theta} e^{i\bar{k}_3z}}{\sqrt{2\bar{k}_3(\bar{k}_0^2-k_\perp^2\cos^{2}\bar{\theta})}},
\end{equation}
the reflected ordinary wave
\begin{equation}\label{ordinary_-}
    \tilde{a}^{(1)}_{\pm}=\frac{\pm i\bar{k}_{3} e^{-i\bar{k}_{3}z\pm i\bar{\theta}}}{\sqrt{2\bar{k}_3(\bar{k}_0^2-k_\perp^2\cos^{2}\bar{\theta})}},\qquad \tilde{A}_3^{(1)}=-\frac{k_\perp \sin\bar{\theta} e^{-i\bar{k}_3z}}{\sqrt{2\bar{k}_3(\bar{k}_0^2-k_\perp^2\cos^{2}\bar{\theta})}},
\end{equation}
the direct extraordinary wave arising from birefringence in a CLC
\begin{equation}\label{extraordinary_+}
\begin{gathered}
        a^{(2)}_{\pm}=\frac{\bar{k}_{3}^{2}\cos\bar{\theta} \pm i\bar{k}_0^{2}\sin\bar{\theta}}{\sqrt{2k_3^{(2)}\bar{k}_0^2(\bar{k}_0^2-k_\perp^2\cos^{2}\bar{\theta})}} e^{iS(\bar{\theta})},\qquad A_3^{(2)}=-\frac{k_3^{(2)} k_\perp\cos\bar{\theta}}{\sqrt{2 k_3^{(2)}\bar{k}_0^2(\bar{k}_0^2-k_\perp^2\cos^{2}\bar{\theta})}} e^{iS(\bar{\theta})},
\end{gathered}
\end{equation}
where
\begin{equation}\label{k32}
     k_3^{(2)}=\sqrt{(1+\de\e)\bar{k}_3^2+\de\e k_\perp^2\sin^{2}\bar{\theta}},\qquad S(\bar{\theta}):=\sqrt{1+\de\e}\frac{\bar{k}_3}{q}E\Big(\bar{\theta};\frac{-\de\e k_\perp^2}{(1+\de\e)\bar{k}_3^2}\Big),
\end{equation}
and the reflected extraordinary wave
\begin{equation}\label{extraordinary_-}
\begin{gathered}
        \tilde{a}^{(2)}_{\pm}=\frac{\bar{k}_{3}^{2}\cos\bar{\theta} \pm i\bar{k}_0^{2}\sin\bar{\theta}}{\sqrt{2k_3^{(2)}\bar{k}_0^2(\bar{k}_0^2-k_\perp^2\cos^{2}\bar{\theta})}} e^{-iS(\bar{\theta})},\qquad \tilde{A}_3^{(2)}=\frac{k_3^{(2)} k_\perp\cos\bar{\theta}}{\sqrt{2 k_3^{(2)}\bar{k}_0^2(\bar{k}_0^2-k_\perp^2\cos^{2}\bar{\theta})}} e^{-iS(\bar{\theta})}.
\end{gathered}
\end{equation}
We have introduce the standard notation,
\begin{equation}
    E(\bar{\theta};\vk):=\int_{0}^{\bar{\theta}}{dx \sqrt{1-\vk\sin^2 x}},
\end{equation}
for the incomplete elliptic integral of the second kind. In calculating the third component, $A_3$, formula \eqref{a_3_comp} has been used, where, within the accuracy we work, one should take into account only the leading in $k_0^{-1}$ term stemming from the action of $\partial_3$ on the fast oscillating exponent. The solutions for the reflected waves \eqref{ordinary_-} and \eqref{extraordinary_-} can be obtained from the solutions \eqref{ordinary_+} and \eqref{extraordinary_+} by using the symmetry,
\begin{equation}\label{symm_transf}
    a_\pm(z)\rightarrow a^*_\mp(z),
\end{equation}
of Eq. \eqref{Max_eqns2}.

In the paraxial regime when the component $A_3$ can be neglected, the polarizations of the ordinary and extraordinary waves are linear. In the case of the ordinary direct and reflected waves, the plane of linear polarization lies at an angle of
\begin{equation}
    \phi=qz
\end{equation}
to the axis $1$. As for the plane of linear polarization of the direct and reflected extraordinary waves, this angle is written as
\begin{equation}
    \phi=\vf+\arg(\bar{k}_3^2\cos\bar{\theta}+i\bar{k}_0^2\sin\bar{\theta})\approx qz.
\end{equation}
In other words, the polarizations vectors of these waves are directed along $\bs\tau(z)$.

Notice that the expressions for the mode functions $\mathbf{A}^{(1,2)}$ and $\mathbf{\tilde{A}}^{(1,2)}$ agree with the symmetry property mentioned in Sec. 5.D of \cite{BKL5} devoted to helical media, the particular case of which the cholesterics are. Namely, on performing the Fourier transform with respect to $\vf=\arg k_{+}$, as in passing from the mode functions of plane-wave photons to the twisted ones \eqref{pln_waves_to_twisted}, the resulting mode functions have the form given in formulas (129) and (131) of \cite{BKL5}. This is a consequence of the fact that the dependence of the mode functions \eqref{ordinary_+}, \eqref{ordinary_-}, \eqref{extraordinary_+}, and \eqref{extraordinary_-} on $z$ is gathered into the combination $\bar{\theta}=qz-\vf$ up to a constant phase factor at the given mode. However, as we shall see further, this property is violated by the presence of interfaces of the CLC plate which spoil the helical symmetry of the system.

The Hamilton-Jacobi action $S(\bar{\theta})$ for the extraordinary wave \eqref{k32} is convenient to split into the linear in $\bar{\theta}$ part and the periodic one, $\bar{S}(\bar{\theta})$, with the period $\pi$:
\begin{equation}
    S(\bar{\theta})=:p_3 \bar{\theta}/q+\bar{S}(\bar{\theta}),
\end{equation}
where the quasimomentum
\begin{equation}
    p_3=\frac{2}{\pi}\sqrt{1+\de\e}\,\bar{k}_3 E\Big(\frac{-\de\e k_\perp^2}{(1+\de\e)\bar{k}_3^2}\Big),\qquad E(x):=E(\pi/2;x).
\end{equation}
It is assumed for the extraordinary waves that the radicand of $k_3^{(2)}$ is nonnegative for any $\bar{\theta}$, i.e., there are no turning points. The presence of turning points means that there exist bound states of photons in the CLC arising due to total internal reflection. In the papers \cite{AksKryuRom06,band,BelOrl91,Shiyan90}, these states were thoroughly investigated and it was shown that they leak from the CLC plate only in the case when $\min(\e_\perp,\e_\parallel)<1$. In what follows, we will consider such values of the momentum $\spk$ of radiated photons that the turning points are not realized.

\begin{figure}[t]
   \centering
\begin{tabular}{cc}
\includegraphics*[width=0.47\linewidth]{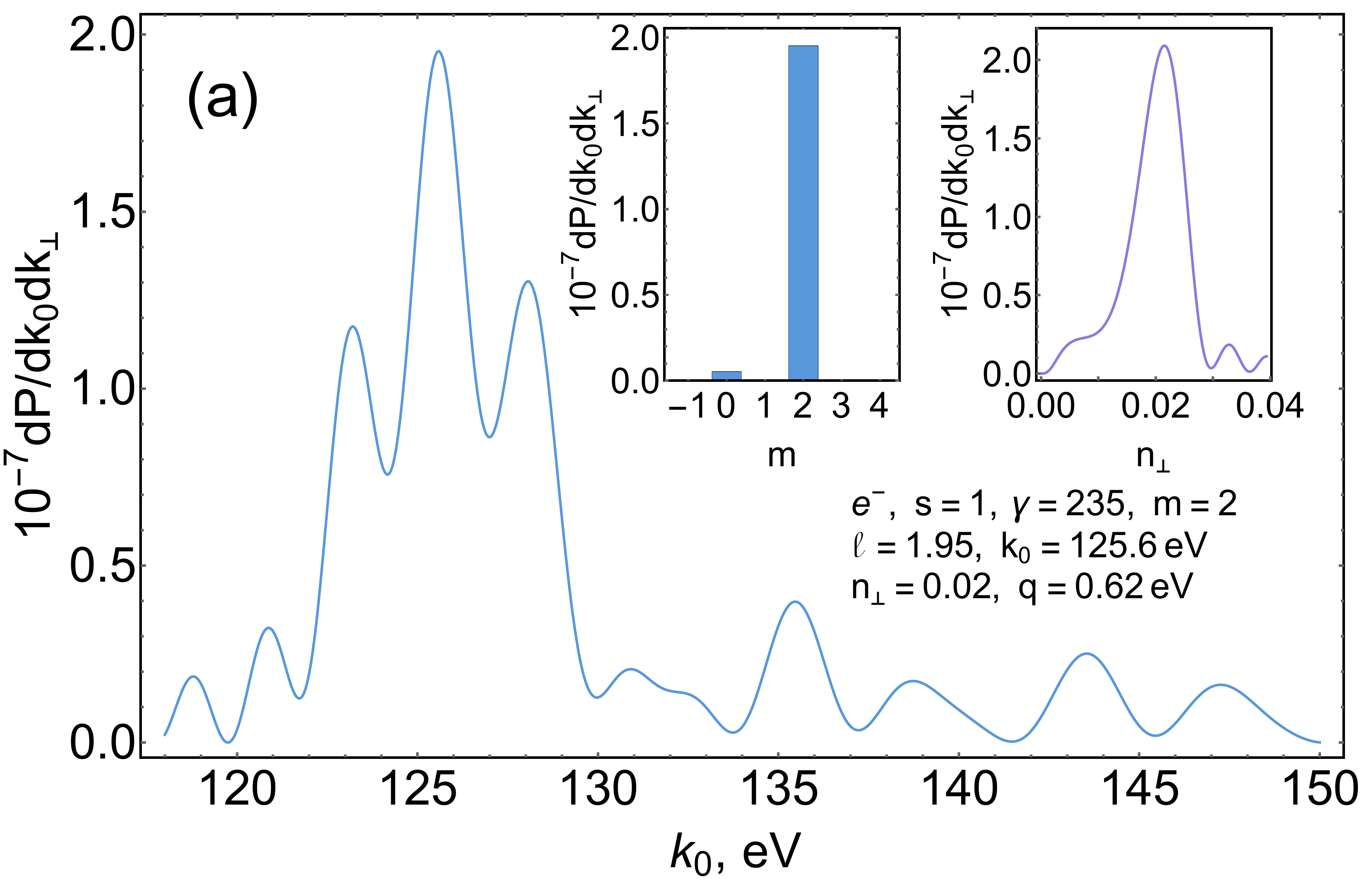}&
\includegraphics*[width=0.47 \linewidth]{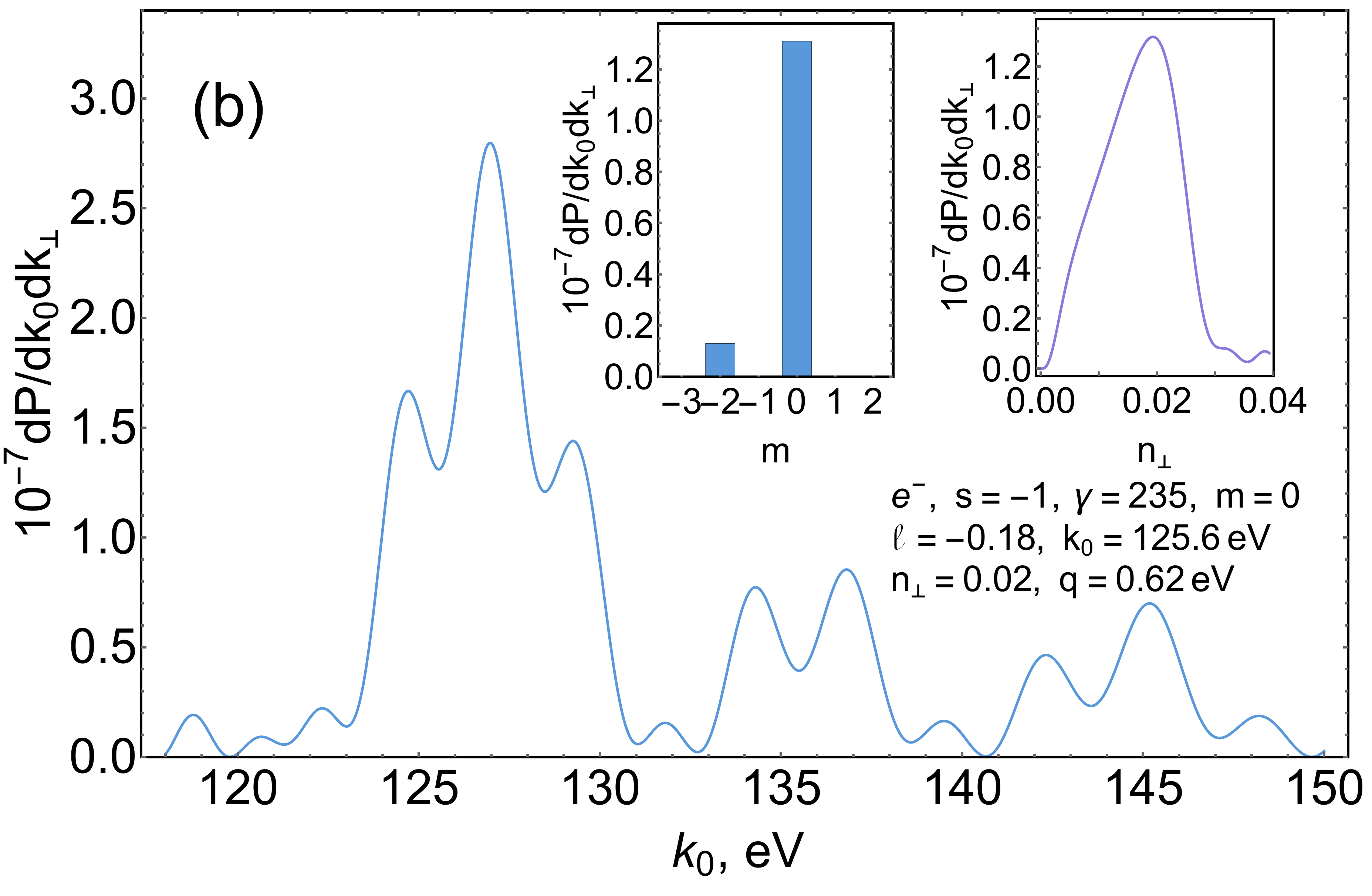}\\
\includegraphics*[width=0.47\linewidth]{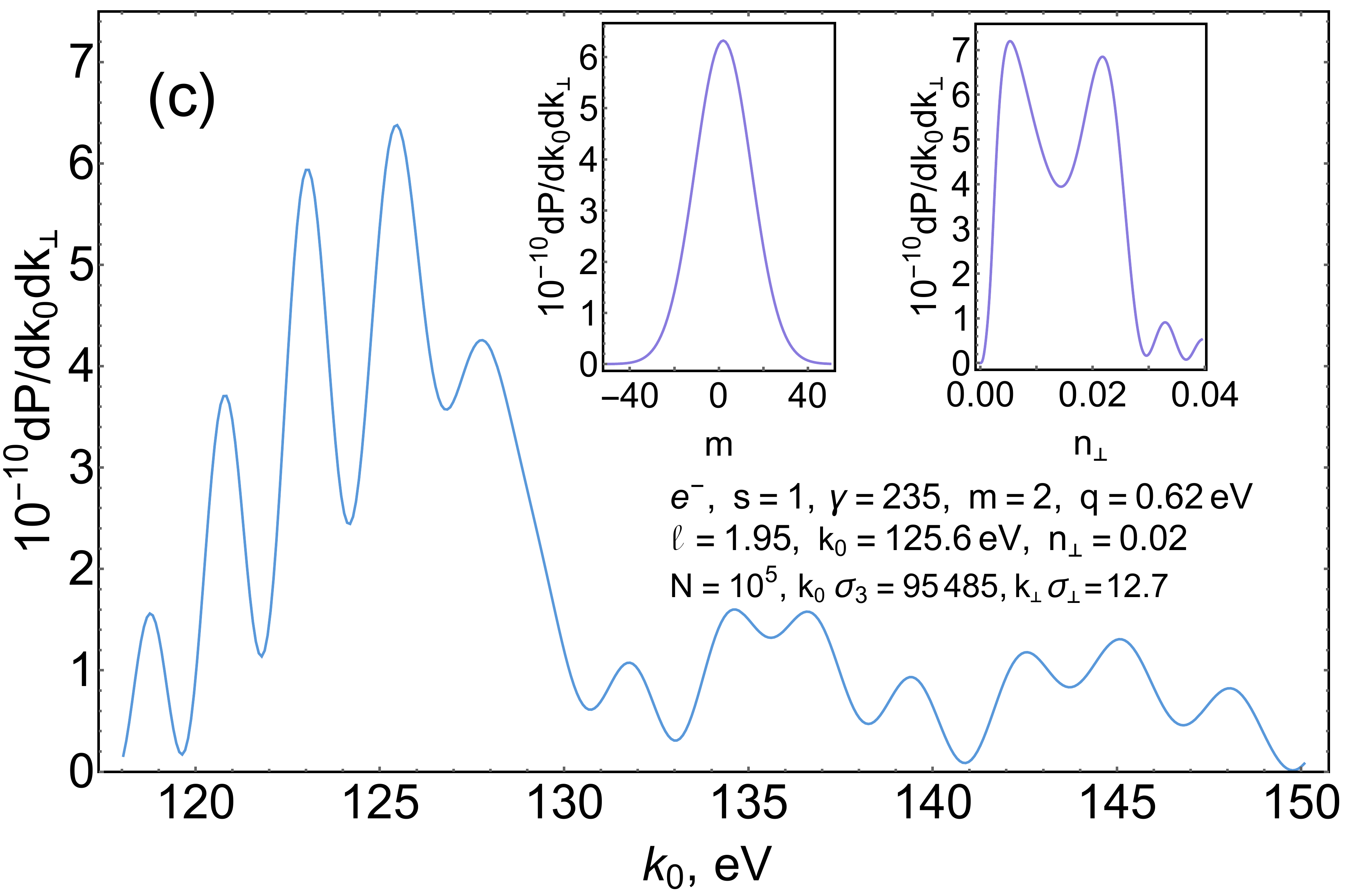}&
\includegraphics*[width=0.47\linewidth]{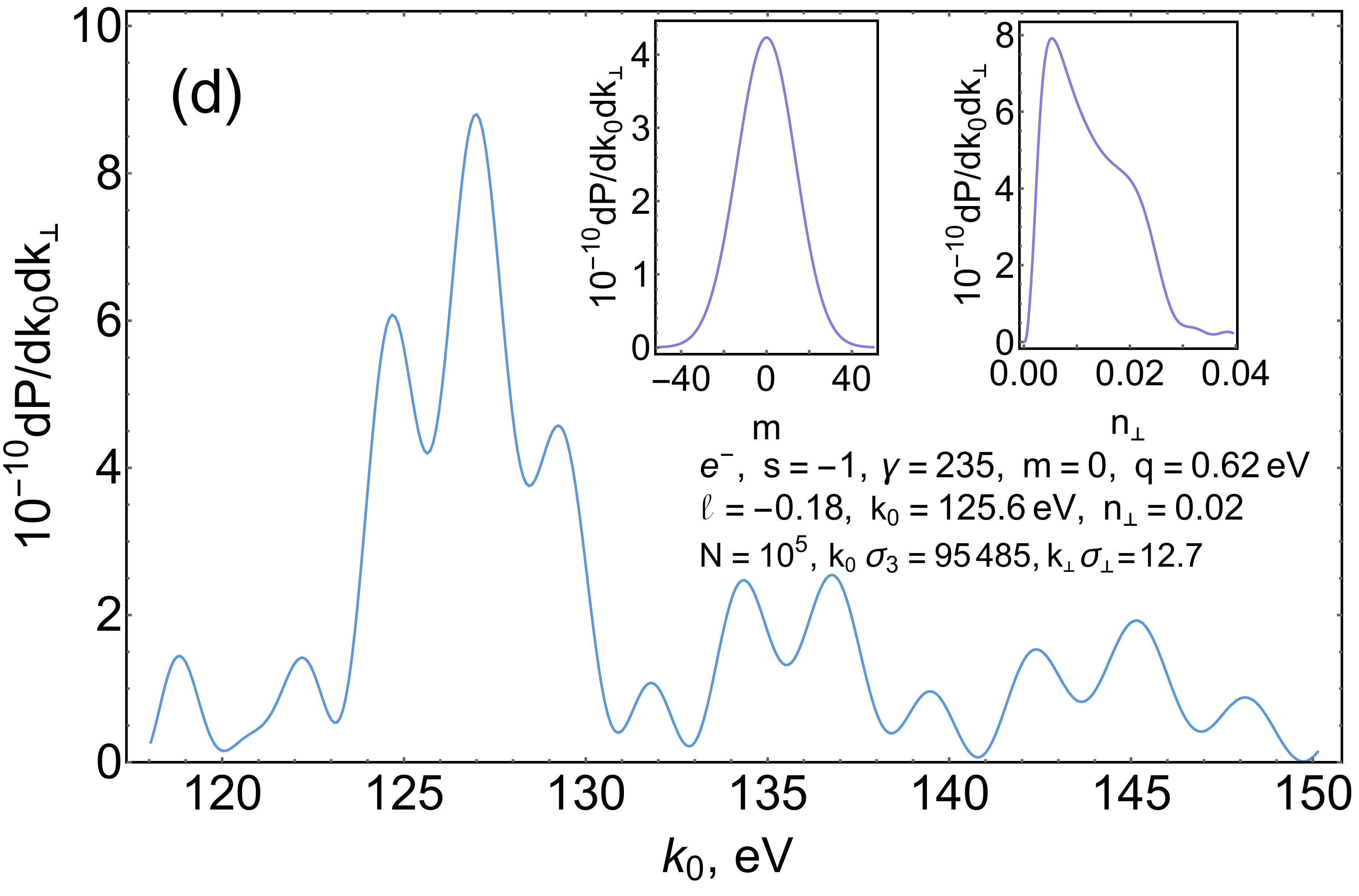}\\
\end{tabular}
    \caption{{\footnotesize The average number of twisted photons, $dP/dk_0dk_\perp$, produced in transition radiation from electrons traversing normally the cholesteric plate. The shortwave approximation is used. The insets on the plots represent the distributions over $m$ and $n_\perp$ at the maximum of radiation intensity. The parameters are the same as in Fig. \ref{eU_wkb_plane_plots}. Upper plots: Transition radiation from the single electron. The plot (a) is for $s=1$ and the plot (b) is for $s=-1$. The selection rule \eqref{selection_rule} is fulfilled for $n=1$ with good accuracy. The contributions with $s=1$ and $s=-1$ are of the same order of magnitude and so the radiation polarization is close to linear. Lower plots: Transition radiation from the Gaussian beam of $N=10^5$ electrons with transverse and longitudinal dimensions $\s_\perp=1$ $\mu$m and $\s_3=125$ $\mu$m, respectively. The plot (c) is for $s=1$ and the plot (d) is for $s=-1$. The distribution over $m$ becomes wider than for the radiation from a single electron since $k_\perp\s_\perp\gg1$. The projection of the total angular momentum per photon, $\ell$, and the projection of the orbital angular momentum per photon are the same as for the one-particle radiation.}}
\label{e235_wkb_plots}
\end{figure}

Consider the CLC plate in a vacuum. This slab is perpendicular to the axis $z$ with the width $L=\pi N_u/q$, where $N_u$ is the number of periods of the CLC helix. Then we have to join the solutions \eqref{Max_eqns} of the Maxwell equations in the CLC plate with the free waves in a vacuum using the standard boundary conditions at $z=0$ and $z=-L$:
\begin{equation}\label{bound_conds_chol}
    [\Phi_\pm]_{z=0}=[\Phi_\pm]_{z=-L}=0,\qquad [\pm i(\partial_3 \Phi_\pm -\partial_\pm \Phi_3)]_{z=0}=[\pm i(\partial_3 \Phi_\pm -\partial_\pm \Phi_3)]_{z=-L}=0.
\end{equation}
For $z>0$, the mode function $\mathbf{\Phi}(s,\spk;x)$ of the quantum electromagnetic field $\hat{\mathbf{A}}(x)$ reads
\begin{equation}\label{mode_funcz>0}
    \mathbf{\Phi}(s,\spk;x)=\frac{C}{\sqrt{2k_0V}}\mathbf{f}(s,\spk)e^{-ik_0x^0+i\spk\spx},
\end{equation}
where the polarization vector of the plane-wave photon with helicity $s$ is written as
\begin{equation}
  \mathbf{f}(s,\spk)=(\cos\vf\cos\theta-is\sin\vf,\sin\vf\cos\theta+is\cos\vf,-\sin\theta)/\sqrt{2},\qquad \sin\theta:=k_\perp/k_0\equiv n_\perp,
\end{equation}
in the Cartesian basis. In the CLC plate, the mode function is given by the linear combination
\begin{equation}\label{mode_func_cho}
  \mathbf{\Phi}(s,\spk;x)=\frac{C}{\sqrt{2k_0V}}(r_1 \mathbf{A}^{(1)}+r_2 \mathbf{A}^{(2)}+l_1 \mathbf{\tilde{A}}^{(1)}+l_2\mathbf{\tilde{A}}^{(2)})e^{-ik_0x^0+i\spk_\perp \spx_\perp},
\end{equation}
where $C$, $r_{1,2}$, and $l_{1,2}$ are some constants and the components of $\mathbf{A}^{(1,2)}$ and $\mathbf{\tilde{A}}^{(1,2)}$ are expressed through \eqref{ordinary_+}, \eqref{ordinary_-}, \eqref{extraordinary_+}, and \eqref{extraordinary_-}. For $z<-L$, the mode function becomes
\begin{equation}\label{mode_funcz<mL}
    \mathbf{\Phi}(s,\spk;x)=\frac{C}{\sqrt{2k_0V}}\big[(d_+\mathbf{f}_{++} +d_-\mathbf{f}_{-+})e^{ik_3z} +(h_+\mathbf{f}_{+-} +h_-\mathbf{f}_{--})e^{-ik_3z}\big]e^{-ik_0x^0+i\spk_\perp\spx_\perp},
\end{equation}
where
\begin{equation}
    \mathbf{f}_{s\s}:=\mathbf{f}(s,\spk_\perp,\s k_3),\qquad\s=\pm1.
\end{equation}
The general formulas for the joining coefficients are given in Appendix \ref{A}.

The final expression for the mode function $\mathbf{\Phi}(s,\spk;x)$ should be normalized. This specifies the normalization constant $C$. The normalization condition gives rise to (see for details Sec. 5.A of \cite{BKL5})
\begin{equation}\label{norm_const0}
    |C|^2=\big(|d_+|^2+|d_-|^2\big)^{-1}.
\end{equation}
It is convenient to write the normalization constant in the form
\begin{equation}\label{norm_const}
    |C|^{-2}=\nu_0+\big[\nu_1 e^{-i(\bar{k}_3+p_3)L} +\nu_2 e^{-2i\bar{k}_3L} +\nu_3 e^{-2ip_3L} +\nu_4 e^{i(\bar{k}_3-p_3)L} +c.c.\big].
\end{equation}
The exact expressions for the coefficients $\nu_{k}$, $k=\overline{0,4}$, and the joining coefficients $r_{1,2}$ and $l_{1,2}$ are rather cumbersome and are given in Appendix \ref{Join_Coeff_App}. In the paraxial approximation, $n^2_\perp\ll1$, the expressions are greatly simplified (see Appendix \ref{C}). We shall show below that it is in the paraxial regime where the radiation from a charged particle moving along the $z$ axis is a pure source of twisted photons.

Having obtained the mode functions, the quantum electromagnetic field $\hat{\mathbf{A}}(x)$ and quantum electrodynamics (QED) in a medium are constructed in the standard way as it is described in \cite{parax,BKL5,GinzbThPhAstr}. The quantum electromagnetic field in the interaction picture is written as
\begin{equation}\label{quantum_field}
    \hat{\mathbf{A}}(x)=\sum_{s=\pm1}\int\frac{V d\spk}{(2\pi)^3}\bs{\Phi}(s,\spk;x)\hat{a}(s,\spk)+\sum_{\al} \bs{\Phi}_\al(x)\hat{a}_\al+\text{H.c.},
\end{equation}
where $\hat{a}$, $\hat{a}^\dag$ denote the creation-annihilation operators
\begin{equation}
    [\hat{a}(s,\spk),\hat{a}^\dag(s',\spk')]=\frac{(2\pi)^3}{V}\de_{s,s'}\de(\spk-\spk'),\qquad [\hat{a}_\al,\hat{a}^\dag_\be]=\de_{\al\be}.
\end{equation}
The second term on the right-hand side of \eqref{quantum_field} describes the contribution of bound states with $n_\perp>1$. This term can be omitted in the analysis of radiation detected out of the CLC plate. The explicit expression for the quantum electromagnetic field allows one to find the probabilities of various QED processes evolving in the CLC slab or near it by employing the standard techniques of QED. In particular, the average number of plane-wave and twisted photons radiated by charged point particles moving along arbitrary trajectories can easily be found \cite{BKL2,BKL5} and the formalism developed in the papers \cite{BKb,BKLb} can be applied. Of course, by using the quantum electromagnetic current one can find the probabilities of other processes involving, for example, Dirac particles \cite{SokLosk57,Riazanov57,GrimNuef95,IvSerZay16,trasrad20} or atomic and nuclear phototransitions \cite{AfSeSol18,atomic,SGACSSK,AfaCarMuck16,SSSF15}.

\section{Radiation of plane-wave photons}\label{PPP}

The peculiarity of the periodic structure of CLCs is revealed in the VC \cite{Orlov,Shipov1,Shipov2,BelVC,Shipov3,CarlosVC,parax} and transition \cite{CarlosTR,parax} radiations. As for any periodic dispersive medium, several Cherenkov cones are produced by a charged particle passing through a CLC plate \cite{Shipov3,BelBook,BazylZhev,Ginzburg}. Because of the constructive interference of electromagnetic waves produced by a charged particle, this radiation possesses the harmonics in energy and has larger intensity at these harmonics than in the case of a homogeneous medium.

We begin with the description of plane-wave photons created by a classical current of the charged particle with the charge $Ze$ moving uniformly and rectilinearly. The trajectory of such a particle reads
\begin{equation}\label{trajectory}
    x^0=t,\qquad \spx=\bs{\be}t.
\end{equation}
The average number of plane-wave photons radiated by a charged particle during an infinite interval of time is written as \cite{GinzbThPhAstr,Ginzburg,LandLifQED,Glaub2,BKL5}
\begin{equation}\label{dP_plane}
    dP(s,\spk)=Z^2e^2\Big|\int_{-\infty}^\infty dt e^{-i k_{0} x^{0}(t)}\dot{\spx}(t)\mathbf{\Phi}(s,\spk;x(t))\Big|^2 \frac{Vd\spk}{(2\pi)^3}.
\end{equation}
The expression on the right-hand side of \eqref{dP_plane} also enters into the probability to detect a photon with the quantum numbers $(s,\spk)$ (see the details in \cite{BKL2,BKL5,epjc}). The expression under the modulus sign, which is proportional to the one-particle transition amplitude, splits into the sum of several terms: the edge radiation from the part of the trajectory $z>0$,
\begin{equation}\label{edge_rad>0}
    \frac{C}{\sqrt{2k_0V}}\frac{i\bs{\beta}\mathbf{f}(s,\spk)}{k_0(1-\mathbf{n}\bs{\beta})};
\end{equation}
the transition radiation from the part of the trajectory $z<-L$,
\begin{equation}\label{edge_rad<mL}
    \frac{iC}{\sqrt{2k_0V}}\Big[\frac{\bs{\be}(\mathbf{f}_{++}d_+ +\mathbf{f}_{+-}d_-)}{k_0(1-\mathbf{n}\bs{\be})}e^{ik_0(1-\mathbf{n}\bs{\be})L/\be_3} +\frac{\bs{\be}(\mathbf{f}_{-+}h_+ +\mathbf{f}_{--}h_-)}{k_0(1-\mathbf{n}_\perp\bs{\be}_\perp+n_3\be_3)}e^{ik_0(1-\mathbf{n}_\perp\bs{\be}_\perp+n_3\be_3)L/\be_3} \Big];
\end{equation}
the transition radiation from the periodic permittivity in the CLC plate
\begin{equation}\label{chol_rad}
\begin{split}
    &\frac{C}{\sqrt{2k_0V}}\sum_{n=-\infty}^\infty e^{-i(2n+1)\vf}\bigg\{ \vf(x^{(1)}_n) r_1 \Big(\frac{\bar{n}_3}{8k_0}\Big)^{1/2} \big[\bar{\beta}_-c_n -\bar{\beta}_+ c_{n+1}-\frac{\be_3n_\perp}{\bar{n}_3}(c_n-c_{n+1}) \big]+\\
    &+\frac{\vf(x^{(2)}_n) r_2 e^{ip_3\vf/q}}{(8\e_\perp k_0)^{1/2}}  \Big[\Big(\e_\perp-\frac{n_\perp^2}{2}\Big)  (\bar{\be}_+ d_{n+1}+\bar{\be}_- d_n) -\frac{n_\perp^2}{2}(\bar{\be}_+ d_{n} +\bar{\be}_- d_{n+1}) -\be_3n_\perp(\tilde{d}_n+\tilde{d}_{n+1})\Big]+\\
    &+\vf(\tilde{x}^{(1)}_n) l_1 \Big(\frac{\bar{n}_3}{8k_0}\Big)^{1/2} \big[\bar{\beta}_-c_n -\bar{\beta}_+ c_{n+1} +\frac{\be_3n_\perp}{\bar{n}_3}(c_n-c_{n+1}) \big]+\\
    &+\frac{\vf(\tilde{x}^{(2)}_n) l_2 e^{-ip_3\vf/q}}{(8\e_\perp k_0)^{1/2}} \Big[\Big(\e_\perp-\frac{n_\perp^2}{2}\Big)  (\bar{\be}_+ d_{-n-1} +\bar{\be}_- d_{-n}) -\frac{n_\perp^2}{2}(\bar{\be}_+ d_{-n} +\bar{\be}_- d_{-n-1}) +\be_3n_\perp(\tilde{d}_{-n}+\tilde{d}_{-n-1})\Big]
    \bigg\},
\end{split}
\end{equation}
where $\bar{\be}_\pm:=\be_\pm e^{\mp i\vf}$, $\be_\perp=|\be_+|$, and
\begin{equation}
    \vf(x):=2\pi e^{iT_uN_ux/2}\de_{N_u}(x),\qquad \de_{N_u}(x):=\frac{\sin(T_uN_ux/2)}{\pi x},\qquad T_u:=\pi/(q\be_3).
\end{equation}
Besides
\begin{equation}\label{x12}
\begin{aligned}
    x^{(1)}_n&:=k_0(1-\bar{n}_3\be_3-\mathbf{n}_\perp\bs{\be}_\perp)-q\be_3(2n+1),&\qquad x^{(2)}_n&:=k_0(1-\bar{n}_3^{(2)}\be_3-\mathbf{n}_\perp\bs{\be}_\perp)-q\be_3(2n+1),\\
    \tilde{x}^{(1)}_n&:=k_0(1+\bar{n}_3\be_3-\mathbf{n}_\perp\bs{\be}_\perp)-q\be_3(2n+1),&
    \qquad \tilde{x}^{(2)}_n&:=k_0(1+\bar{n}_3^{(2)}\be_3-\mathbf{n}_\perp\bs{\be}_\perp)-q\be_3(2n+1),\\
    \bar{n}_3&:=\bar{k}_3/k_0,&\qquad \bar{n}_3^{(2)}&:=p_3/k_0.
\end{aligned}
\end{equation}
In order to obtain \eqref{chol_rad}, we have expanded the periodic part of the integrand of \eqref{dP_plane} in a Fourier series and introduced the notation for the Fourier coefficients
\begin{equation}\label{cn_dn}
\begin{split}
    c_n&:=\int_{-\pi}^\pi\frac{d x}{2\pi}\frac{e^{-2in x}}{(\e_\perp-n_\perp^2\cos^2x)^{1/2}}=\frac{(2|n|-1)!!}{\e_\perp^{1/2}|n|!} \Big(\frac{n_\perp^2}{8\e_\perp}\Big)^{|n|}F(|n|+1/2,|n|+1/2;2|n|+1;n_\perp^2/\e_\perp),\\
    d_n&:=\int_{-\pi}^\pi\frac{d x}{2\pi}\frac{e^{-2in x+i\bar{S}(x)}}{(\e_\parallel -n_\perp^2(1+\de\e\cos^2x))^{1/4}(\e_\perp-n_\perp^2\cos^2x)^{1/2}},\\
    \tilde{d}_n&:=\int_{-\pi}^\pi\frac{d x}{2\pi} e^{-2in x+i\bar{S}(x)} \frac{(\e_\parallel -n_\perp^2(1+\de\e\cos^2x))^{1/4}}{(\e_\perp-n_\perp^2\cos^2x)^{1/2}}.
\end{split}
\end{equation}
The integral defining the coefficients $c_n$ can be found in \cite{PrBrMar1}. In the expression on the last line of \eqref{chol_rad}, we have used the oddness of the function $\bar{S}(\bar{\theta})$.

In the paraxial limit, the last two integrals in \eqref{cn_dn} are readily evaluated. Namely, if the following estimates are satisfied,
\begin{equation}\label{paraxial_cond2}
    n_\perp^2\ll1,\qquad \frac{\e_\parallel^{1/2}k_0}{32q}\Big(\frac{n_\perp^2\de\e}{\e_\parallel}\Big)^2\ll1,
\end{equation}
then
\begin{equation}
    \bar{S}(\bar{\theta})=\sqrt{1+\de\e}\frac{\bar{k}_3}{q}\Big[E\Big(\bar{\theta};\frac{-n_\perp^2\de\e}{(1+\de\e)\bar{n}_3^2}\Big) -\frac{2}{\pi}E\Big(\frac{-n_\perp^2\de\e}{(1+\de\e)\bar{n}_3^2}\Big)\Big]\approx-\frac{k_\perp n_\perp\de\e}{8q\e_\parallel^{1/2}}\sin(2\bar{\theta}).
\end{equation}
Therefore, we have
\begin{equation}\label{dn_appr}
    d_n\approx\e_\parallel^{-1/4}\e_\perp^{-1/2} J_{-n}\Bigg(\frac{k_\perp n_\perp\de\e}{8q\e_\parallel^{1/2}}\Bigg),\qquad \tilde{d}_n\approx \e_\parallel^{1/4}\e_\perp^{-1/2} J_{-n}\Bigg(\frac{k_\perp n_\perp\de\e}{8q\e_\parallel^{1/2}}\Bigg).
\end{equation}
Furthermore, in the leading order in $n_\perp$, we obtain
\begin{equation}\label{cn_appr}
    c_n\approx\frac{(2|n|-1)!!}{\e_\perp^{1/2}|n|!}\Big(\frac{n_\perp^2}{\e_\perp}\Big)^{|n|}.
\end{equation}
For $n_\perp\rightarrow0$, the coefficients of the Fourier series $c_n$, $d_n$, and $\tilde{d}_n$ are proportional to $\de_{n,0}$. In that case, the series over $n$ in \eqref{chol_rad} is terminated.

\begin{figure}[t]
   \centering
\begin{tabular}{cc}
\includegraphics*[width=0.47\linewidth]{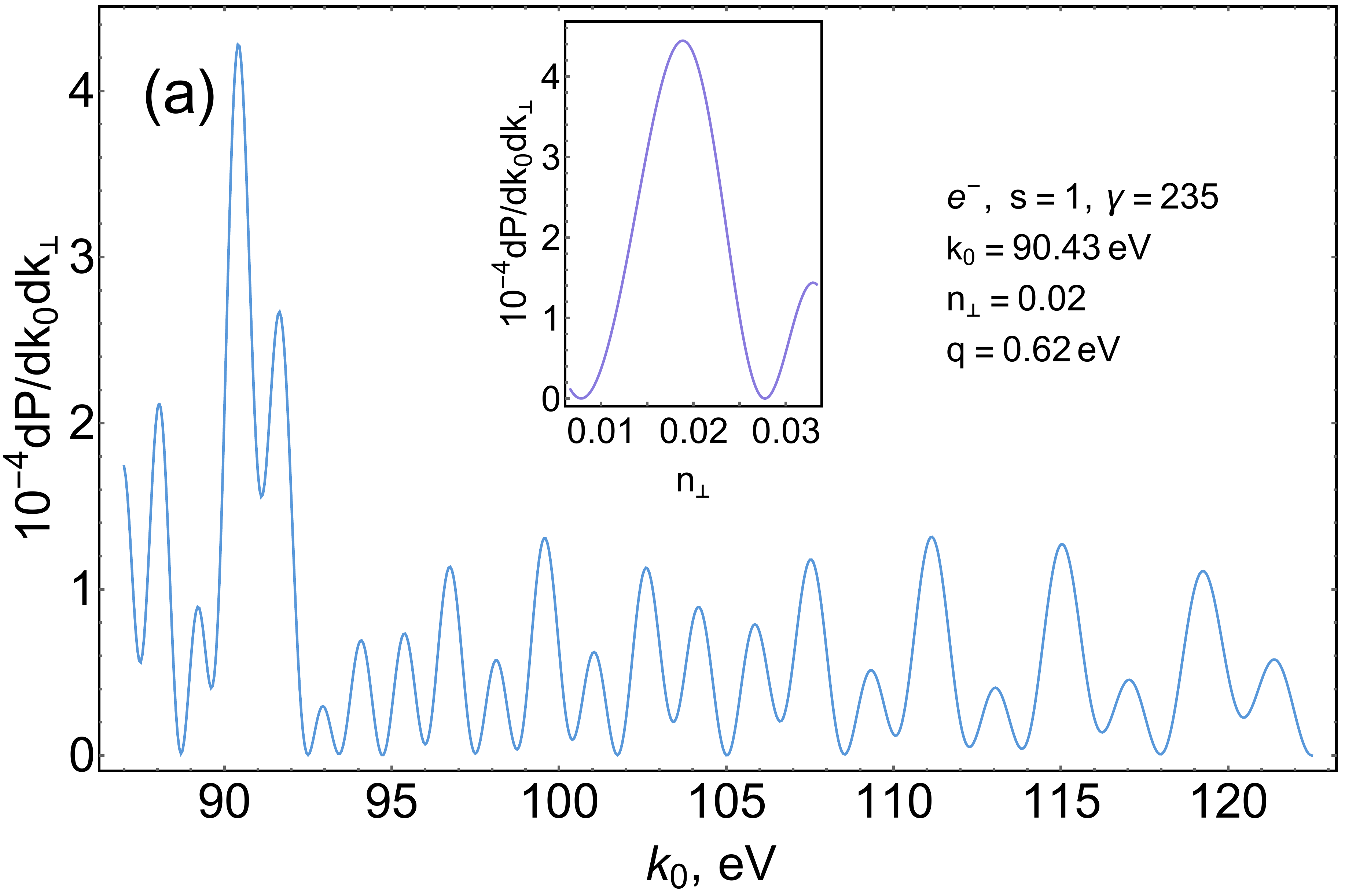}&
\includegraphics*[width=0.47 \linewidth]{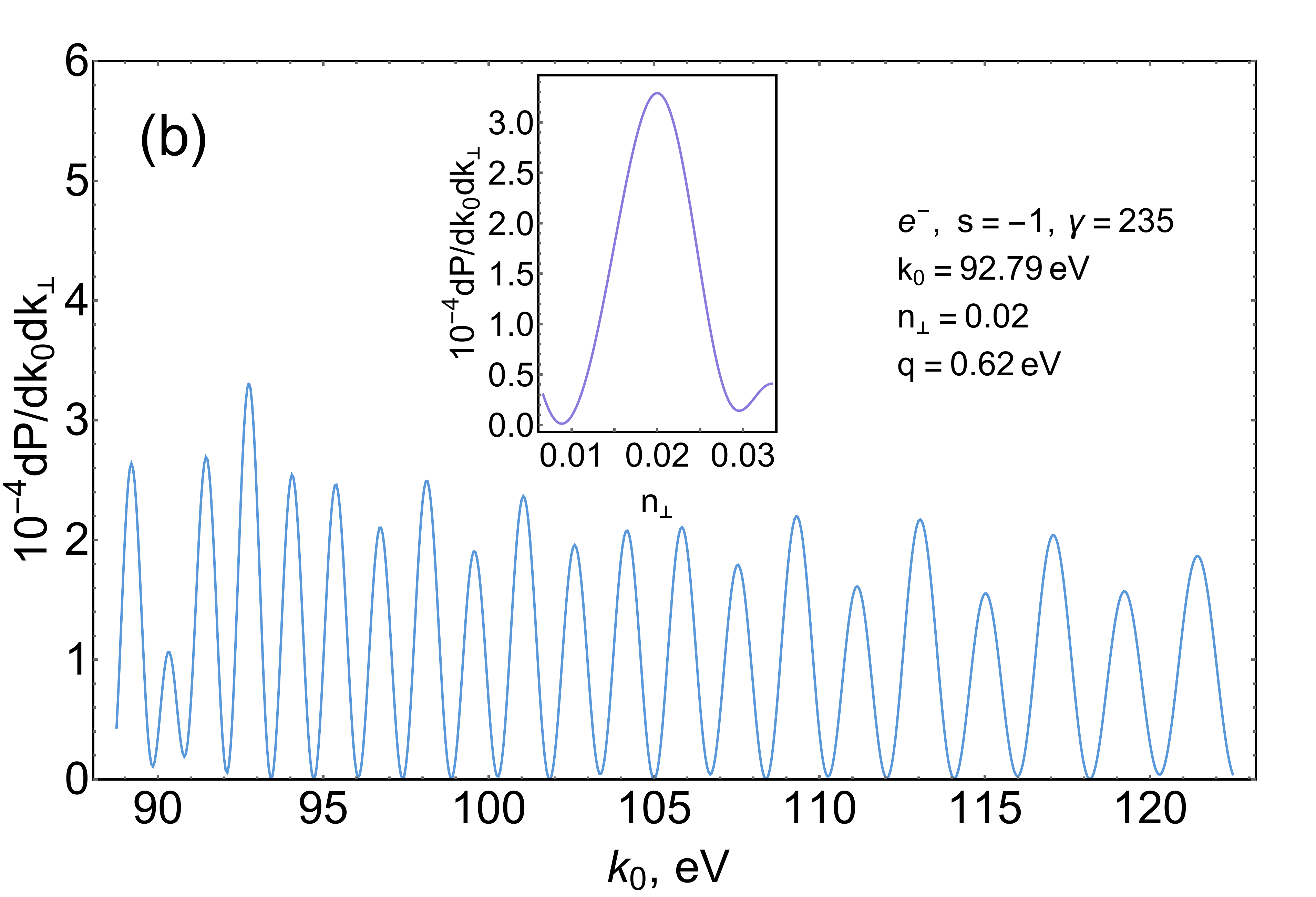}\\
\includegraphics*[width=0.47\linewidth]{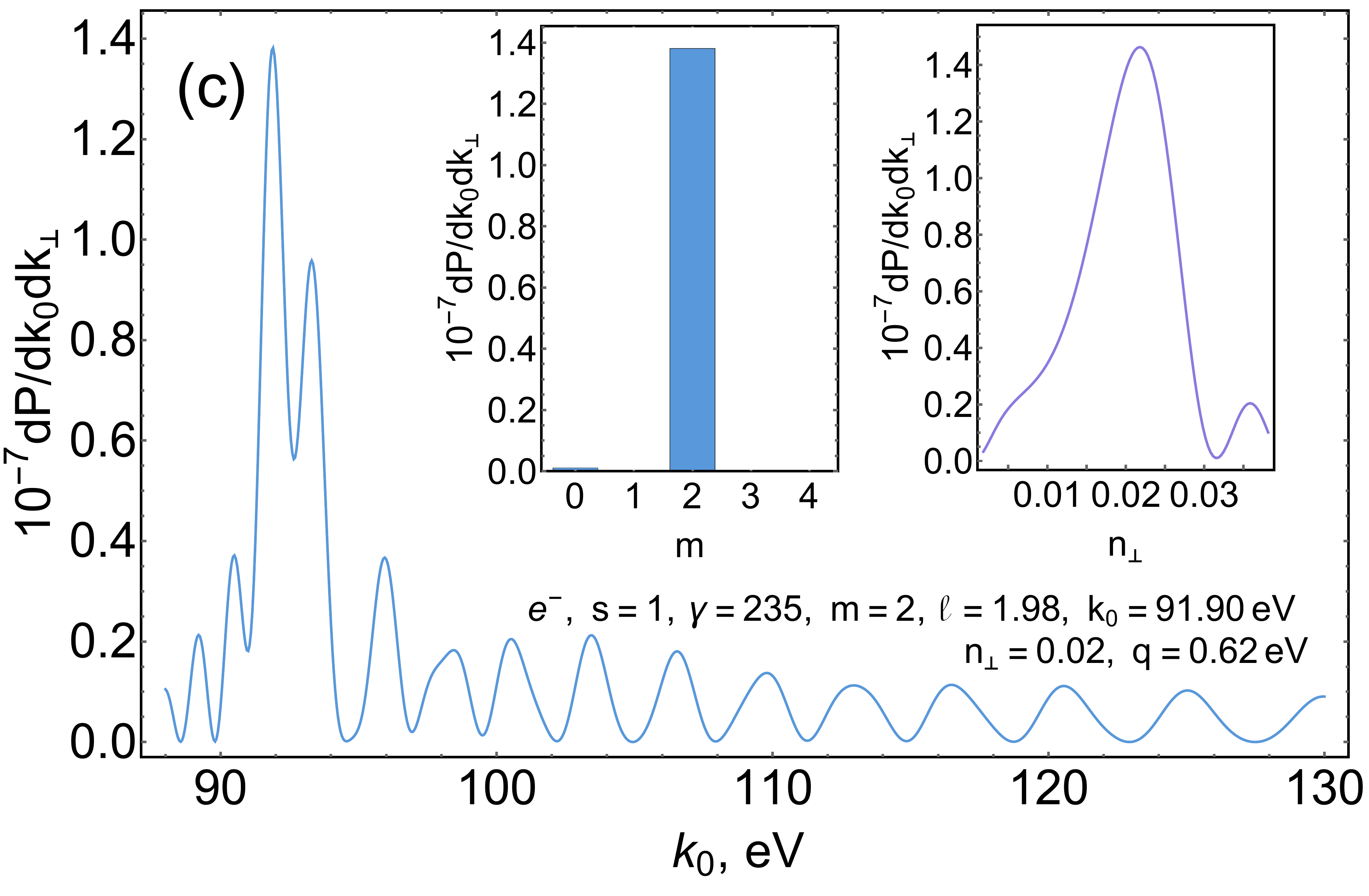}&
\includegraphics*[width=0.47\linewidth]{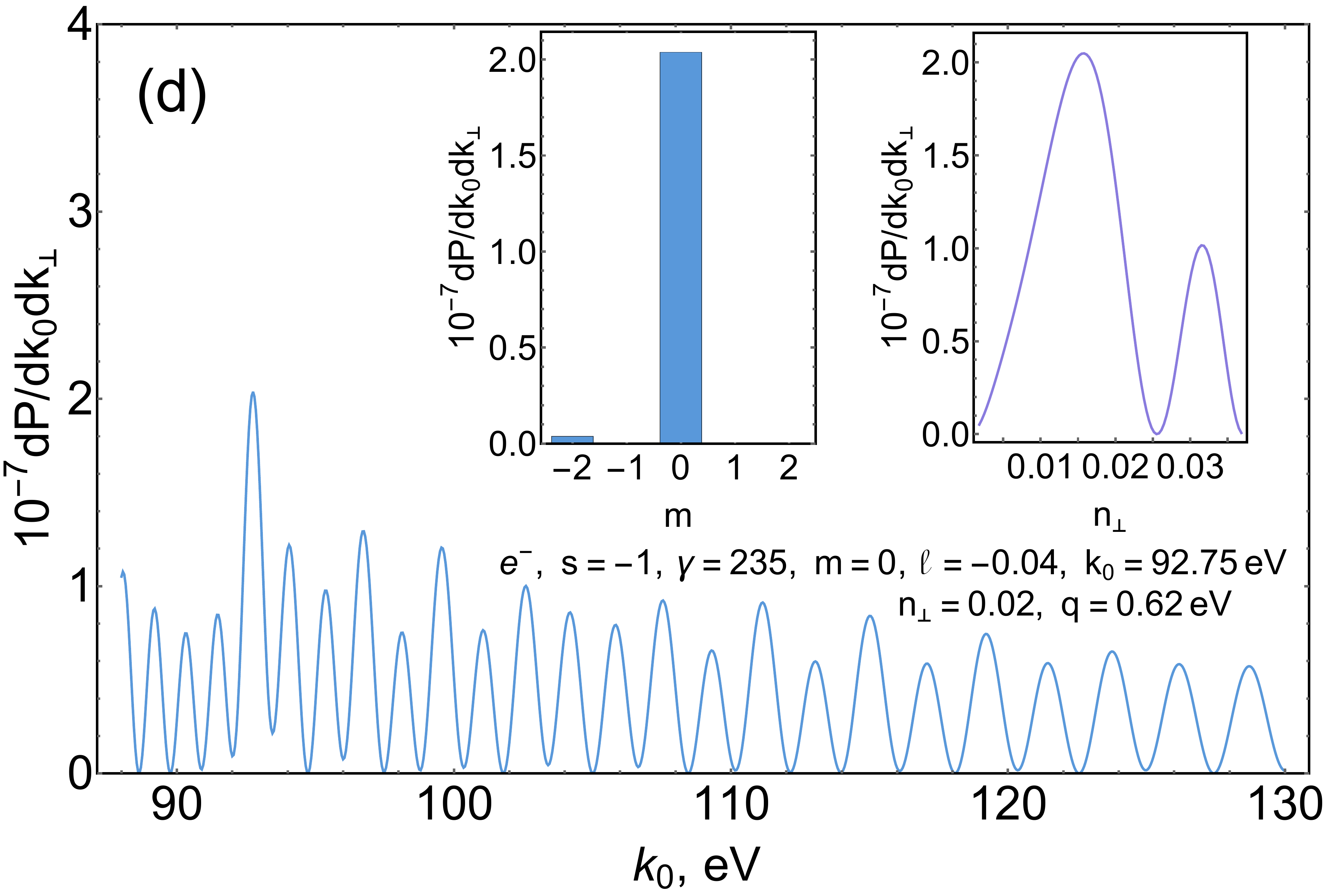}\\
\end{tabular}
    \caption{{\footnotesize The same as on the lower plots in Fig. \ref{eU_wkb_plane_plots} and on the upper plots in Fig. \ref{e235_wkb_plots} but the paraxial approximation is employed \cite{parax}. The insets on the plots represent the distributions over $m$ and $n_\perp$ at the maximum of radiation intensity. For the parameters taken, the paraxial approximation is applicable (the smallness parameter following from \eqref{applic_conds_parax} is $2.1\times 10^{-3}$) whereas the shortwave approximation (the parameter \eqref{short_wave2} is $1.7$) and the perturbation theory with respect to $\de\e$ (the smallness parameter following from \eqref{appl_cond_de} is $3.2\times 10^{3}$ ) are not. Despite this fact we see that the shortwave approximation describes the distributions with respect to $m$ and $n_\perp$ quite well. The distributions over $k_0$ agree only qualitatively. The selection rule \eqref{selection_rule} holds for both the approximations. Upper plots: Transition radiation of plane-wave photons. The plot (a) is for $s=1$ and the plot (b) is for $s=-1$. Lower plots: Transition radiation of twisted photons. The plot (c) is for $s=1$ and the plot (d) is for $s=-1$.}}
\label{e235_parax_wkb_plots}
\end{figure}

We shall be interested further in the case $N_u\gg1$. Then the modulus of the function $\vf(x)$ possesses a sharp maximum at $x=0$. Therefore, squaring the modulus of the one-particle amplitude, one can neglect the contributions of the edge radiation \eqref{edge_rad>0}, \eqref{edge_rad<mL} and of the interference terms resulting from the product of the terms in \eqref{chol_rad} standing at $\vf(x)$ with different arguments. Hence
\begin{equation}\label{ProbPho}
\begin{split}
    &dP(s,\spk)\approx |ZeC|^2 \frac{d\spk}{32\pi k_0}\sum_{n=-\infty}^\infty  \bigg\{ \de_{N_u}^2(x^{(1)}_n) |r_1|^2 \frac{\bar{n}_3^2}{k_0} \Big|\bar{\beta}_-c_n -\bar{\beta}_+ c_{n+1}-\frac{\be_3n_\perp}{\bar{n}_3}(c_n-c_{n+1}) \Big|^2+\\
    &+\de_{N_u}^2(x^{(2)}_n)\frac{|r_2|^2}{\e_\perp k_0} \Big|\Big(\e_\perp-\frac{n_\perp^2}{2}\Big)  (\bar{\be}_+ d_{n+1}+\bar{\be}_- d_n) -\frac{n_\perp^2}{2}(\bar{\be}_+ d_{n} +\bar{\be}_- d_{n+1}) -\be_3n_\perp(\tilde{d}_n+\tilde{d}_{n+1})\Big|^2+\\
    &+\de_{N_u}^2(\tilde{x}^{(1)}_n) |l_1|^2 \frac{\bar{n}_3^2}{k_0} \Big|\bar{\beta}_-c_n -\bar{\beta}_+ c_{n+1} +\frac{\be_3n_\perp}{\bar{n}_3}(c_n-c_{n+1}) \Big|^2+\\
    &+\de_{N_u}^2(\tilde{x}^{(2)}_n)\frac{|l_2|^2}{\e_\perp k_0} \Big|\Big(\e_\perp-\frac{n_\perp^2}{2}\Big)  (\bar{\be}_+ d_{-n-1} +\bar{\be}_- d_{-n}) -\frac{n_\perp^2}{2}(\bar{\be}_+ d_{-n} +\bar{\be}_- d_{-n-1}) +\be_3n_\perp(\tilde{d}_{-n}+\tilde{d}_{-n-1})\Big|^2
    \bigg\} .
\end{split}
\end{equation}
The energy spectrum of radiation is found from the requirement that the argument of $\de_{N_u}^2(x)$ vanishes. As a result, formulas \eqref{x12} imply the four spectral series:
\begin{equation}\label{spectrum}
\begin{aligned}
    k_0&=\frac{q\be_3(2n+1)}{1-\mathbf{n}_\perp\bs{\be}_\perp-\bar{n}_3\be_3},&\qquad k_0&=\frac{q\be_3(2n+1)}{1-\mathbf{n}_\perp\bs{\be}_\perp+\bar{n}_3\be_3},\\
    k_0&=\frac{q\be_3(2n+1)}{1-\mathbf{n}_\perp\bs{\be}_\perp-\bar{n}^{(2)}_3\be_3},&\qquad k_0&=\frac{q\be_3(2n+1)}{1-\mathbf{n}_\perp\bs{\be}_\perp+\bar{n}^{(2)}_3\be_3},
\end{aligned}
\end{equation}
where $n\in \mathbb{Z}$ and $k_0>0$. In the case of CLC with small electric susceptibility, $\chi_{\parallel,\perp}:=\e_{\parallel,\perp}-1$, in the paraxial approximation, the approximate radiation spectrum produced by an ultrarelativistic particle, $\ga\gg1$, $\be_\perp\ll1$, takes a simple form
\begin{equation}\label{spectrum_ap}
\begin{aligned}
    k_0&=\frac{2q\ga^2(2n+1)}{1+(\bs{\be}_\perp-\mathbf{n}_\perp)^2\ga^2-\chi_\perp\ga^2},&\qquad k_0&=\frac{q(2n+1)}{2+\chi_\perp/2},\\
    k_0&=\frac{2q\ga^2(2n+1)}{1+(\bs{\be}_\perp-\mathbf{n}_\perp)^2\ga^2-\chi_\parallel\ga^2},&\qquad k_0&=\frac{q(2n+1)}{2+\chi_\parallel/2},
\end{aligned}
\end{equation}
As we see, the shortwave approximation \eqref{short_wave1}, \eqref{short_wave2} is not applicable for description of radiation induced by the quantum field modes corresponding to the reflected waves \eqref{ordinary_-}, \eqref{extraordinary_-} when the harmonic number $n$ is small.

For $\chi_{\parallel,\perp}\ga^2>1$, there are values of $n_\perp$ such that the denominator in formulas \eqref{spectrum} for the spectrum of radiation induced by the direct ordinary and extraordinary waves is negative. In that case, $q(2n+1)$ must be negative. In the opposite case, the quantity $q(2n+1)$ must be positive. Notice that all the above formulas and properties are valid for both right-handed, $q>0$, and left-handed, $q<0$, CLCs. The schematic representation of the possible experimental setup for observation of the created radiation is given in Fig. \ref{scheme_plots}. The plots of the average number of plane-wave photons produced by the uranium nuclei and electrons crossing normally the CLC plate are presented in Fig. \ref{eU_wkb_plane_plots}.

\begin{figure}[t]
   \centering
\begin{tabular}{cc}
\includegraphics*[width=0.47\linewidth]{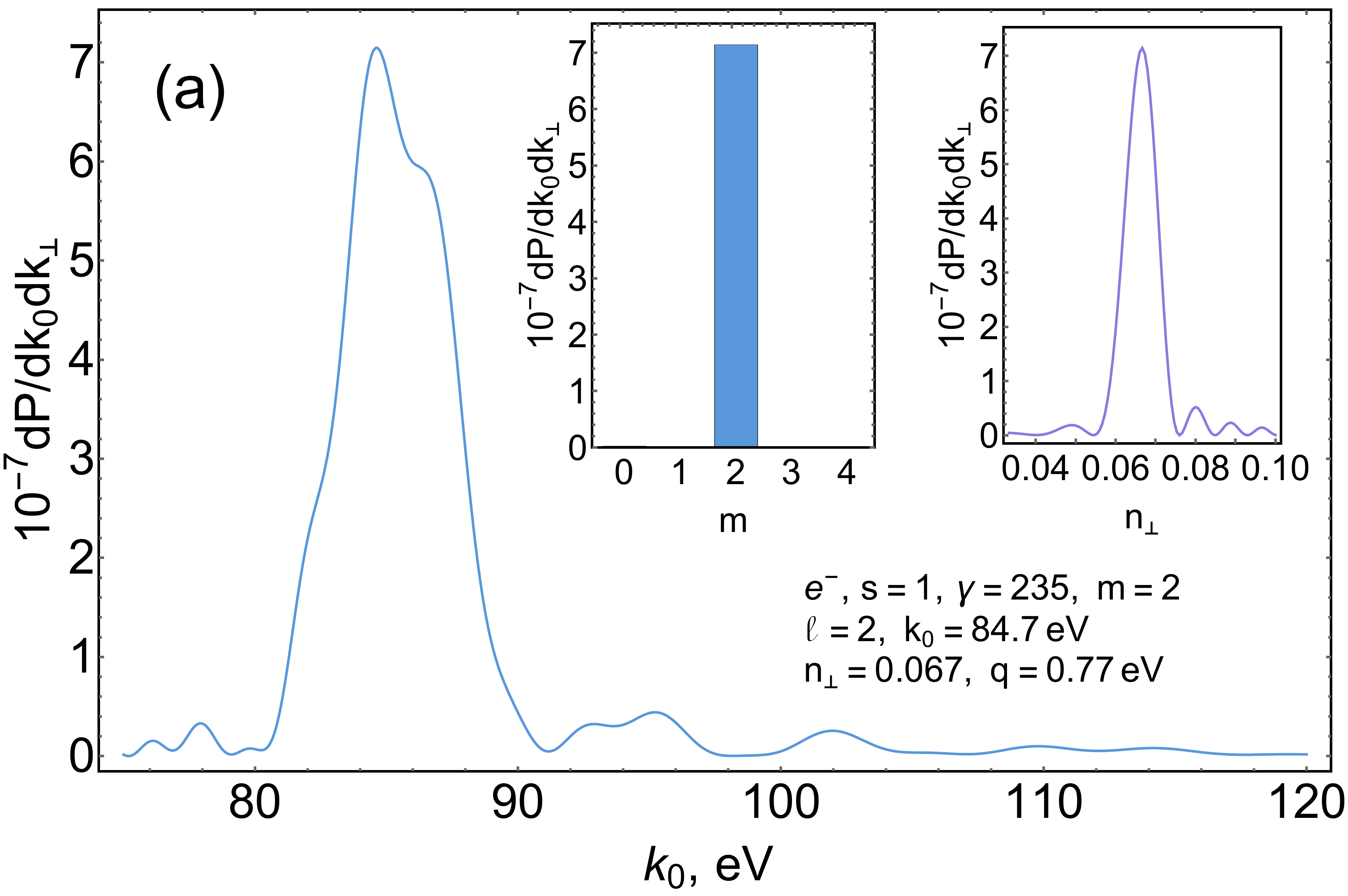}&
\includegraphics*[width=0.47 \linewidth]{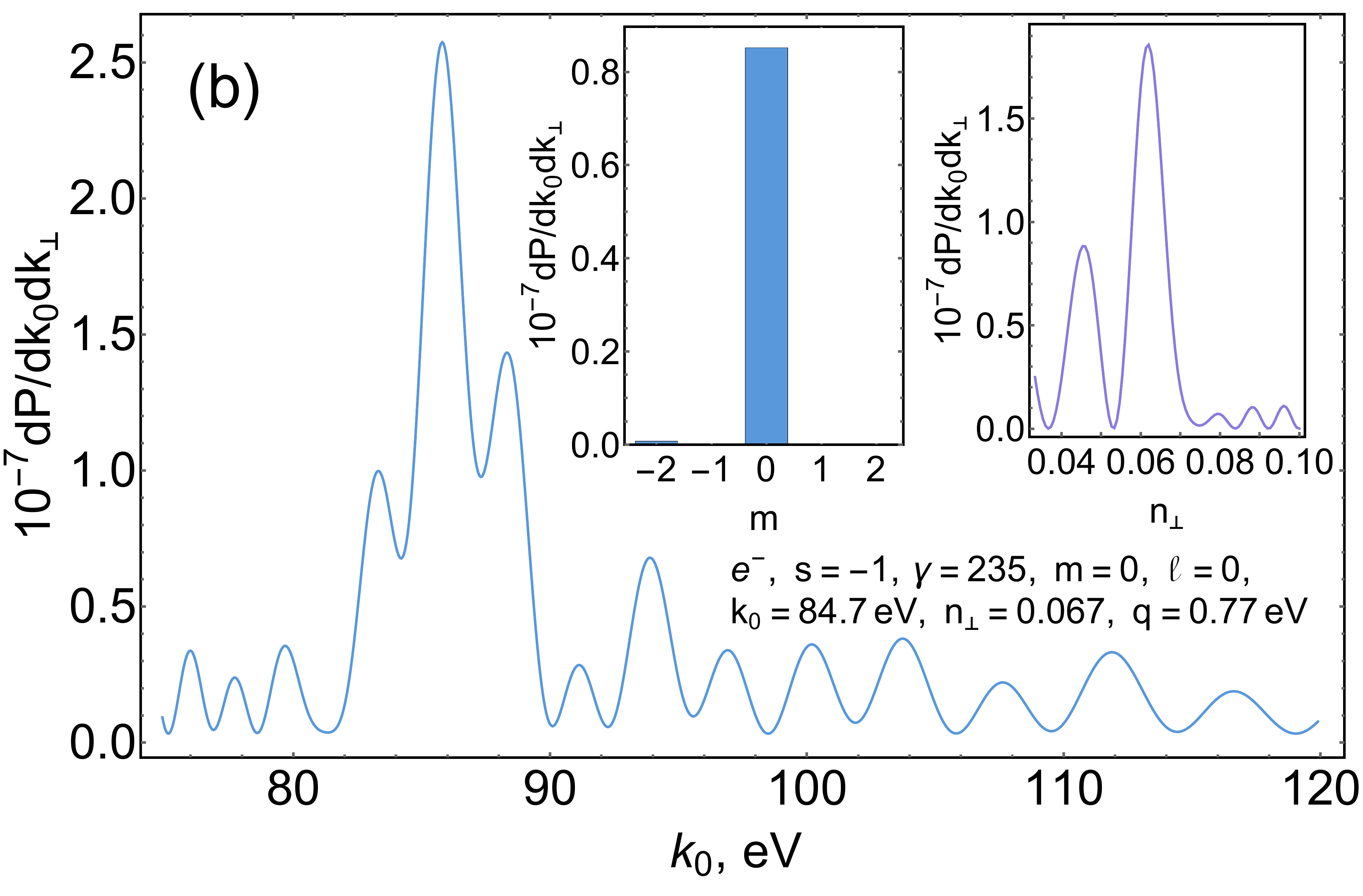}\\
\includegraphics*[width=0.47\linewidth]{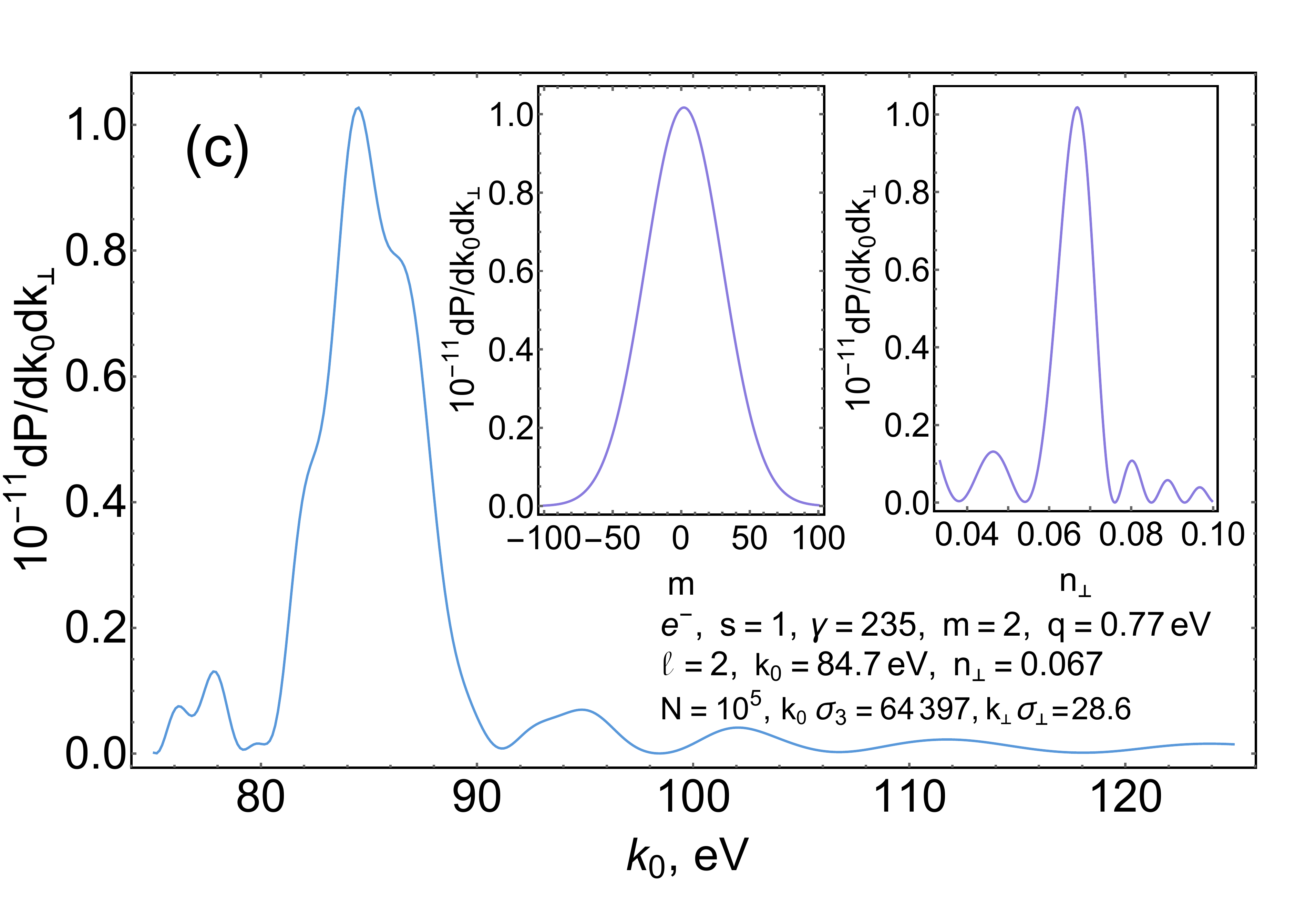}&
\includegraphics*[width=0.47\linewidth]{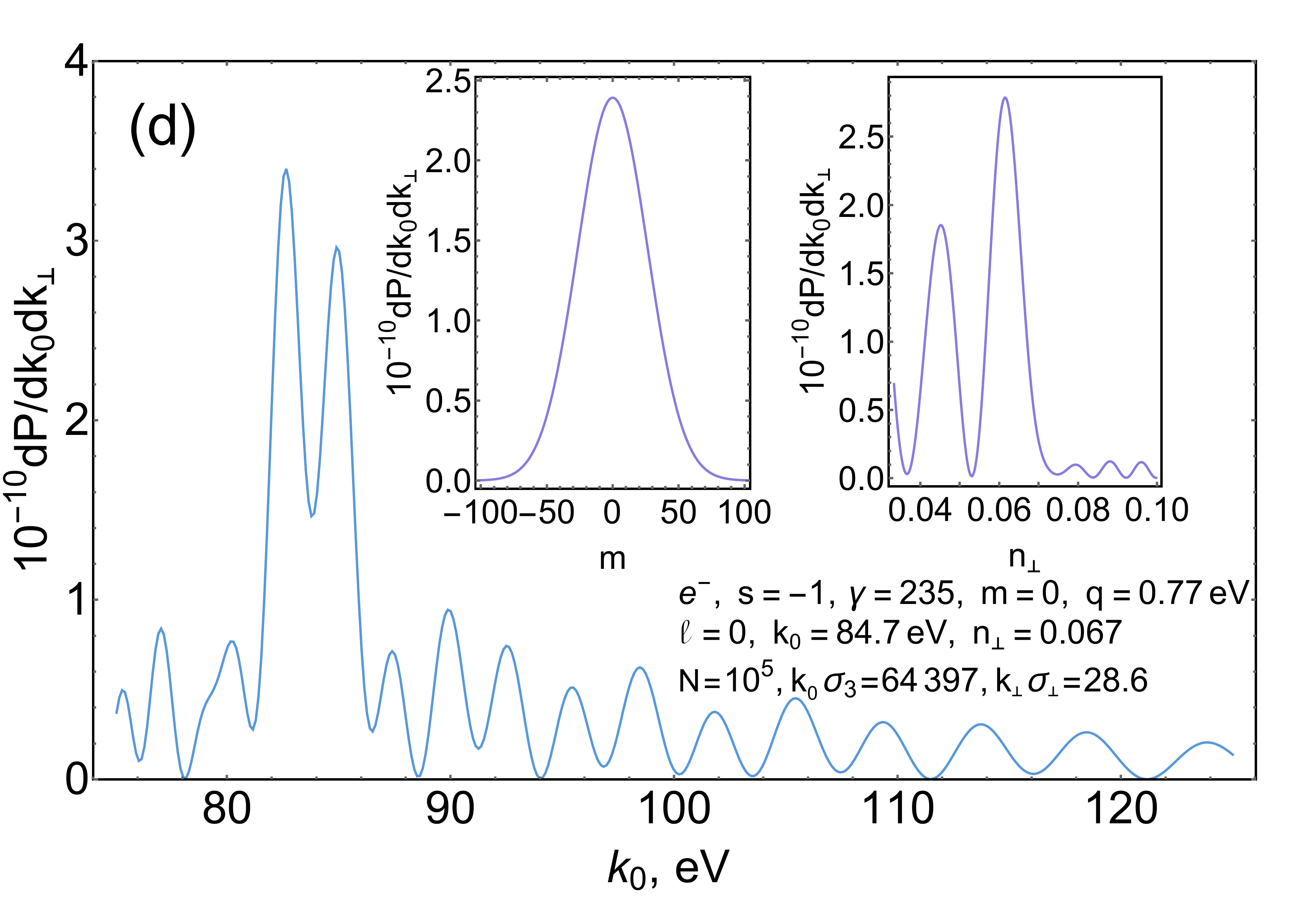}\\
\end{tabular}
    \caption{{\footnotesize The average number of twisted photons, $dP/dk_0dk_\perp$, produced in transition radiation from electrons traversing normally the cholesteric plate. The paraxial approximation is used \cite{parax}. The insets on the plots represent the distributions over $m$ and $n_\perp$ at the maximum of radiation intensity. The width of the CLC plate $L=20$ $\mu$m, the number of periods $N_u=25$, and the components of the permittivity tensor are $\e_\perp=1-\omega_p^2/(3 k_0^2)$ and $\e_\parallel=1-\omega_p^2/k_0^2$ with $\omega_p=21$ eV. The parameter \eqref{short_wave2} is $1.3$ at the maximum of radiation intensity and so the shortwave approximation does not work well. In particular, we see that the radiation at the harmonic is not linearly polarized. The perturbation theory with respect to $\de\e$ is also not valid for these parameters. The maximum of the applicability parameters \eqref{appl_cond_de} is $2.5\times 10^2$. The maximum of the parameters  \eqref{applic_conds_parax} controlling the applicability of the perturbation theory with respect to $n_\perp$ is $2.0\times10^{-2}$. Upper plots: Transition radiation from a single electron. The plot (a) is for $s=1$ and the plot (b) is for $s=-1$. The selection rule \eqref{selection_rule} is fulfilled for $n=1$. Lower plots: Transition radiation from the Gaussian beam of $N=10^5$ electrons with transverse and longitudinal dimensions $\s_\perp=1$ $\mu$m and $\s_3=125$ $\mu$m, respectively. The plot (c) is for $s=1$ and the plot (d) is for $s=-1$. The distribution over $m$ becomes wider than for the radiation from a single electron since $k_\perp\s_\perp\gg1$. The projection of the total angular momentum per photon, $\ell$, and the projection of the orbital angular momentum per photon are the same as for the one-particle radiation.}}
\label{e235_parax_plots}
\end{figure}

\section{Radiation of twisted photons}\label{ProbTwPhWKB}

Inasmuch as the translations in the plane $(x,y)$ are a symmetry of the CLC permittivity, it was useful to start our study with consideration of radiation of plane-wave photons. Having obtained the one-particle radiation amplitude of plane-wave photons, it is not difficult to find the radiation amplitude of twisted photons. To this aim, one needs to take a linear combination of radiation amplitudes of plane-wave photons as in \cite{SerboNew,JenSerprl,JenSerepj}. The expansion of the plane-wave photon wave function in terms of the twisted ones reads
\begin{equation}\label{pln_waves_to_twisted}
    \frac{\mathbf{f}(s,\spk)e^{i\spk\spx}}{\sqrt{2k_0V}}=-\frac{\sin\theta e^{ik_3z}}{2\sqrt{k_0V}} \sum_{m=-\infty}^\infty i^me^{-im\vf}\sqrt{RL_z} \Big(\frac{2}{\sin\theta}\Big)^{3/2} \bs{\psi}(s,m,k_3,k_\perp;\spx),
\end{equation}
where the components of the mode functions of twisted photons are \cite{SerboNew,JenSerprl,JenSerepj,BKL2}
\begin{equation}
\begin{split}
    \psi_3(s,m,k_3,k_\perp;\spx)&=\frac{1}{\sqrt{RL_z}} \Big(\frac{k_\bot}{2 k_{0}}\Big)^{3/2}J_m(k_\perp|x_+|)e^{im\arg x_+}, \\ \psi_\pm(s,m,k_3,k_\perp;\spx)&=\frac{i k_{\bot}}{s k_{0}\pm k_{3}}\psi_3(s,m\pm1,k_3,k_\perp;\spx).
\end{split}
\end{equation}
The parameters $V$, $R$, and $L_z$ characterize the normalization volume for the system at issue.

In order to find the average number of twisted photons radiated by a charged particle traversing the CLC plate along the trajectory \eqref{trajectory}, it is sufficient to expand the plane-wave photon wave functions entering into \eqref{edge_rad>0}-\eqref{chol_rad} in terms of the twisted photon wave functions employing formula \eqref{pln_waves_to_twisted}. As in the previous section, we shall be interested in the leading contribution to radiation at $N_u\gg1$. Besides, we suppose that $\bs{\be}_\perp=0$ and the estimates \eqref{paraxial_cond2} are valid, i.e., we work in the paraxial approximation. In the paraxial regime, the projection of the total angular momentum $m$ of the twisted photon splits into the spin part $s$ and the projection of the orbital angular momentum $l=m-s$ \cite{OAM}.

Under the above assumptions, one can neglect the contributions proportional to $n_\perp^2$ in the normalization constant \eqref{norm_const} and use the approximate expressions \eqref{ri_li} for the joining coefficients. Applying \eqref{pln_waves_to_twisted} to the expression on the first line in \eqref{chol_rad}, we deduce the following contribution to the one-particle radiation amplitude of a twisted photon with quantum numbers $s$, $m$, $k_3$, and $k_\perp$:
\begin{equation}
    -\frac{s\be_3 C}{4\sqrt{RL_z}}i^{-m}\Big(\frac{n_\perp}{2}\Big)^{3/2}\sum_{n=-\infty}^\infty \vf(x_n)\de_{m,2n+1+s} (1+\e_\perp^{-1/2})(c_n-c_{n+1}).
\end{equation}
The other terms in \eqref{chol_rad} are transformed in the same way. Introducing, for brevity, the notation
\begin{equation}\label{J}
    \tilde{J}_n:=J_n\bigg(\frac{k_{\bot} n_\perp \de\e }{8q\e_\parallel^{1/2}}\bigg),\; n\in \mathbb{Z};\qquad \tilde{J}_n(x):=0,\; n\not\in \mathbb{Z},
\end{equation}
we derive the average number of radiated twisted photons
\begin{equation}\label{ProbTP}
\begin{split}
    dP(s,m,k_3,k_{\perp})=\,&|Ze\be_3 C|^2 \sum_{n=-\infty}^\infty \Big[ \de_{N_u}^2(x^{(1)}_{n})\de_{m,2n+1+s}(1+\e_\perp^{-1/2})^2(c_n-c_{n+1})^2+\\
    &+\de_{N_u}^2(x^{(2)}_{n})\tilde{J}^2_{(m-2n-1-s)/2}\frac{(1+\e_{\parallel}^{1/2})^2}{\e_{\parallel}^{1/2}\e_\perp} (\tilde{d}_n+\tilde{d}_{n+1})^2+\\
    &+\de_{N_u}^2(\tilde{x}^{(1)}_{n})\de_{m,2n+1+s}(1-\e_\perp^{-1/2})^2(c_n-c_{n+1})^2+\\
    &+\de_{N_u}^2(\tilde{x}^{(2)}_{n})\tilde{J}^2_{(m-2n-1-s)/2}\frac{(1-\e_{\parallel}^{1/2})^{2}}{\e_\parallel^{1/2}\e_\perp} (\tilde{d}_n+\tilde{d}_{n+1})^2 \Big]\frac{n_{\bot}^{3} dk_3 dk_\perp}{64},
\end{split}
\end{equation}
where $\tilde{d}_n$ and $c_n$ have the form \eqref{dn_appr}, \eqref{cn_appr}, and $x^{(1,2)}_{n}$, $\tilde{x}^{(1,2)}_{n}$ are given by formulas \eqref{x12} with $\bs{\be}_\perp=0$. The photon energy spectrum has the same form as for the plane-wave photons \eqref{spectrum}, \eqref{spectrum_ap} but with $\bs{\be}_\perp=0$. Since $\tilde{d}_n$ and $c_n$ are proportional to $\de_{n,0}$ in the paraxial limit, the main contribution to radiation comes from the harmonics with $n=\{-1,0\}$. Notice that the average number of radiated twisted photons \eqref{ProbTP} summed over $m$ is independent of $s$. This reflects the fact that the radiation is linearly polarized in the approximation we use.

As long as the photon energy in \eqref{spectrum}, \eqref{spectrum_ap} must be positive, the quantity $q(2n+1)>0$ for radiated twisted photons corresponding to reflected waves. Furthermore, in the ultrarelativistic limit, one can introduce the standard notation for the Cherenkov cone opening
\begin{equation}\label{VCh_cone}
    n_\perp^{\text{VC}}:=\sqrt{\chi_{\parallel,\perp}-\ga^{-2}},
\end{equation}
for the direct ordinary and extraordinary waves. In that case, if $n_\perp^{\text{VC}}$ is imaginary or $n_\perp>n_\perp^{\text{VC}}$, then $q(2n+1)>0$ for the photons radiated in this mode. If $0<n_\perp<n_\perp^{\text{VC}}$, then $q(2n+1)<0$. Hence, in the both cases, only one value of $n$ from $\{-1,0\}$ is admissible for a given $n_\perp$. For $n_\perp^{\text{VC}}>0$, the one value of $n$ from $\{-1,0\}$ is realized at $n_\perp<n_\perp^{\text{VC}}$ whereas the other value of $n$ is realized at $n_\perp>n_\perp^{\text{VC}}$.

As is seen from the expression \eqref{ProbTP}, the photon radiated by a charged particle due to ordinary waves obeys the selection rule
\begin{equation}\label{selection_rule}
    l:=m-s=2n+1.
\end{equation}
This selection rule forbids the radiation of twisted photons with even values of $l$ by a charged particle traversing normally a CLC plate. Moreover, taking into account that the harmonics with $n=\{-1,0\}$ dominate, the twisted photons with $l=\pm1$ are mostly radiated. The twisted photons radiated at a given harmonic are linearly polarized and possess the OAM projection $l$.

It also follows from the above analysis of the spectrum that the handedness of the CLC helix determines the sign of the OAM projection of radiated twisted photons. The sign flip of the CLC helix chirality changes the sign of the OAM projection of radiated photons provided other parameters are not varied. Moreover, if the harmonic of the direct radiation lies inside of the Cherenkov cone \eqref{VCh_cone}, then  $ql<0$ (see Fig. \ref{U_wkb_plots}). If the harmonic of the direct radiation is outside of the Cherenkov cone, then $ql>0$ (see Figs. \ref{e235_wkb_plots}, \ref{e235_parax_wkb_plots}, \ref{e235_parax_plots}). This inversion of the sign of the orbital angular momentum of radiated photons is analogous to the sign flip of $l$ due to the anomalous Doppler effect in undulators filled with dispersive medium \cite{BKL8}. The same properties are inherent to twisted photons radiated by a charged particle due to extraordinary waves when
\begin{equation}\label{J<<1}
   \frac{k_{\bot} n_\perp |\de\e| }{8|q|\e_\parallel^{1/2}}\ll1,
\end{equation}
i.e. in the paraxial regime.

Recall that there is the selection rule $l=-s$ for twisted photons produced by transition and VC radiations from a classical charged particle crossing a plate made of an isotropic homogeneous dielectric \cite{BKL5,BKL7}. In order to observe the twisted photons in those radiations, it is necessary to install a homogeneous circular polarizer selecting the photons with definite helicity. In the case of a CLC plate, such a polarizer is unnecessary. This becomes especially important for the hard twisted photons as long as it is difficult to select the photons with definite helicity in this spectral range.

\section{Conclusion}

Let us summarize the results. We constructed the quantum theory of radiation of plane-wave and twisted photons from classical charged particles traversing a CLC plate. Investigations in this field are conducted for many years \cite{BelBook,Orlov,Shipov1,Shipov2,BelVC,Shipov3,CarlosVC,parax}. However, these studies were mainly based on the two wave approximation or on the perturbation theory with resect to the anisotropy of the permittivity tensor or on the paraxial approximation. In the present paper, we investigated the radiation of twisted and plane-wave photons in the shortwave approximation for which $k_0\gg q$, the observation angles are arbitrary and $\de\e$ can be large.

The shortwave approximation was already used to describe the propagation of electromagnetic waves in liquid crystals \cite{AksValRom01,AksValRom04,AksKryuRom06,Aksenova2008}. This method showed a good agreement with experimental data in the range of its applicability. Employing this method, we obtained the explicit expressions for the mode functions taking into account the boundary conditions on the interfaces of the CLC plate for arbitrary parameters of the radiated photon where the shortwave approximation is valid. This allowed us to construct the operator of a quantum electromagnetic field in the CLC plate. The derived mode functions can also be used for investigation of optical properties of the CLC plate of a finite width and for description of radiation produced by charged particles moving in CLCs along trajectories of a general form.

Having the operator of a quantum electromagnetic field at hand, the explicit expressions for the average numbers of plane-wave \eqref{ProbPho} and twisted \eqref{ProbTP} photons were deduced by the standard means \cite{LandLifQED,BKL5,parax}. As expected, the radiation generated by a charged particle moving uniformly and rectilinearly possesses the harmonics in energy of radiated photons \eqref{spectrum}. In virtue of anisotropy of the CLC permittivity tensor and of the finite width of the plate, there are four series of harmonics induced by the direct ordinary and extraordinary waves and their reflected counterparts. The shortwave approximation proved to be not applicable for description of photon radiation corresponding to reflected waves at low harmonics. One needs to resort to other methods \cite{parax} for a correct description of radiation at these harmonics. Nevertheless the shortwave approximation gives a rather good qualitative description even in the parameter domain where it is poorly applicable (see Fig. \ref{e235_parax_wkb_plots}).

We analyzed the radiation spectrum with respect to the projection of the total and orbital angular momenta at the given harmonics. It turns out that, in the paraxial regime, there is the selection rule $l=2n+1$, where $l$ is the OAM projection of a radiated twisted photon and $n\in \mathbb{Z}$ is the harmonic number. In other words, the system at issue is a pure source of twisted photons. Since the pitch of the CLC helix can be changed easily by varying the temperature, the external electromagnetic field, and the composition of CLC, this source allows one to change quickly the energy of radiated twisted photons in a wide spectral range. In the shortwave approximation, the twisted photons radiated from a classical charged particle traversing normally the CLC plate possess a linear polarization. The peculiar form of the CLC permittivity tensor gives rise to generation of twisted photons mainly with $l=\pm1$ in the paraxial regime at the harmonics $n=\{-1,0\}$ (see Figs. \ref{U_wkb_plots}, \ref{e235_wkb_plots}, \ref{e235_parax_wkb_plots}, \ref{e235_parax_plots}). The radiation at the higher harmonics is suppressed. The radiation harmonics corresponding to $n$ with larger absolute value can be amplified by making use of the coherent radiation of microbunched beams of electrons adjusted to such a frequency of photon radiation. Nowadays, there are created the electron bunch trains with distinguishable harmonics of coherent radiation in the X-ray range \cite{Hemsing7516,PRRibic19}. Furthermore, the distribution of radiation of twisted photons with respect to $m$ can be shifted by the number of the harmonic of coherent radiation \cite{BKLb} if one uses the coherent radiation from helically microbunched beams of charged particles \cite{HemStuXiZh14}.

In comparison with transition radiation from charged particles crossing the plate made of an isotropic homogeneous dielectric or an ideal conductor where the OAM projection $l=-s$, in the case of the CLC plate $l$ does not depend on the photon polarization. This allows one to simplify the experimental scheme proposed in \cite{BKL7} to observe the twisted photons in transition and VC radiations. One needs no a homogeneous circular polarizer in the case of a CLC plate. The source of twisted photons based on the transition radiation from a CLC plate can be used to rotate the micro- and nano-objects in physics, chemistry, and biology and also to excite the nondipole transitions in atoms \cite{SSSF15,AfaCarMuck16,SGACSSK,atomic} and nuclei \cite{SerboNew,AfSeSol18}.

\paragraph{Acknowledgments.}

This work was supported by the Russian Science Foundation (Project No. 17-72-20013).

\appendix
\section{General formulas for the joining coefficients}\label{A}

Let us find the joining coefficients for the mode functions \eqref{mode_funcz>0}, \eqref{mode_func_cho}, and \eqref{mode_funcz<mL} and the normalization factor. The joining coefficients are obtained from the boundary conditions \eqref{bound_conds_chol}. In evaluating the derivatives of the semiclassical mode functions \eqref{ordinary_+}, \eqref{ordinary_-}, \eqref{extraordinary_+}, \eqref{extraordinary_-}, one has to take into account only the leading contributions in powers of $k^{-1}_0$. These contributions come from the derivatives acting on the fast oscillating exponent. It is useful to write the corresponding conditions as the matrix equation
\begin{equation}\label{joining_eqn}
    \left[
       \begin{array}{cc}
         U & 0 \\
         UT & -U_0T' \\
       \end{array}
     \right]
     \left[
       \begin{array}{c}
         a_{ch} \\
         a_l \\
       \end{array}
     \right]=
     \left[
       \begin{array}{c}
         g \\
         0 \\
       \end{array}
     \right],
\end{equation}
where
\begin{equation}\label{ach_def}
    a^T_{ch}=(r_1,r_2,l_1,l_2),\qquad a_l^T=(d_+,d_-,h_+,h_-).
\end{equation}
Besides,
\begin{equation}\label{joining_eqn_3}
\begin{gathered}
    T=\diag(e^{-i\bar{k}_3L},e^{-ip_3L},e^{i\bar{k}_3L},e^{ip_3L}),\qquad    T'=\diag(e^{-ik_3L},e^{-ik_3L},e^{ik_3L},e^{ik_3L}),\\
    U_0=\frac{1}{\sqrt{2}}
    \left[
      \begin{array}{cccc}
        \cos\theta-1 & \cos\theta+1 & -\cos\theta-1 & -\cos\theta+1 \\
        \cos\theta+1 & \cos\theta-1 & -\cos\theta+1 & -\cos\theta-1 \\
        k_0(1-\cos\theta) & k_0(1+\cos\theta) & k_0(1+\cos\theta) & k_0(1-\cos\theta) \\
        k_0(1+\cos\theta) & k_0(1-\cos\theta) & k_0(1-\cos\theta) & k_0(1+\cos\theta) \\
      \end{array}
    \right],\\
    U=
    \left[
      \begin{array}{cc}
        U_{11} & U_{12} \\
        U_{21} & U_{22} \\
      \end{array}
    \right],\qquad
    U_{11}=
    \left[
      \begin{array}{cc}
        a_+^{(1)} & a_+^{(2)} \\
        a_-^{(1)} & a_-^{(2)} \\
      \end{array}
    \right],\qquad U_{12}=U_{11}\Big|_{a\rightarrow \tilde{a}},\\
    U_{21}=-iK \partial_zU_{11},\qquad U_{22}=-iK \partial_zU_{12}=-U_{21}\Big|_{a\rightarrow \tilde{a}},\\
    g^T=\big[\cos\theta-s,\cos\theta+s,k_0(1-s\cos\theta),k_0(1+s\cos\theta)\big]/\sqrt{2},
\end{gathered}
\end{equation}
where $a_\pm^{(1,2)}$ and $\tilde{a}_\pm^{(1,2)}$ are the solutions \eqref{ordinary_+}, \eqref{ordinary_-}, \eqref{extraordinary_+}, and \eqref{extraordinary_-}. In the expressions for the components of the matrix $U$, all the functions of $z$ are taken at $z=0$. Explicitly,
\begin{equation}
    U_{21}=
    \left[
      \begin{array}{cc}
        \bar{k}_3\big[a_+^{(1)}+\frac{k_\perp^2}{2\bar{k}_3^2}(a_+^{(1)}+a_-^{(1)})\big] & k^{(2)}_3\big[a_+^{(2)}+\frac{k_\perp^2}{2\bar{k}_3^2}(a_+^{(2)}+a_-^{(2)})\big] \\
        \bar{k}_3\big[a_-^{(1)}+\frac{k_\perp^2}{2\bar{k}_3^2}(a_-^{(1)}+a_+^{(1)})\big] & k^{(2)}_3\big[a_-^{(2)}+\frac{k_\perp^2}{2\bar{k}_3^2}(a_-^{(2)}+a_+^{(2)})\big] \\
      \end{array}
    \right].
\end{equation}
Then we have
\begin{equation}\label{joining_sol}
    a_{ch}=U^{-1}g,\qquad a_l=(T')^{-1}U_0^{-1}UTU^{-1}g,
\end{equation}
for the solution of Eq. \eqref{joining_eqn}.

Despite the fact that we work in the shortwave approximation, the solution \eqref{joining_sol} obeys exactly the unitarity relation (see for details Appendix A in \cite{parax}):
\begin{equation}
    1+|h_+|^2+|h_-|^2=|d_+|^2+|d_-|^2.
\end{equation}
This allows one to write the normalization constant in the form (see the details in Sec. 5.A of \cite{BKL5})
\begin{equation}
    |C|^2=2(1+a^\dag_l a_l)^{-1}=\big(|d_+|^2+|d_-|^2\big)^{-1}.
\end{equation}

\section{Explicit expressions for the joining coefficients}\label{Join_Coeff_App}

The exact expressions for the joining coefficients $a_{ch}$ defined in \eqref{ach_def} can be cast into the form
\begin{equation}\label{ach_wkb}
\begin{split}
    r_1&=is\sqrt{\frac{k_0}{\e_\perp-n_\perp^2\cos^2\vf}}(\bar{n}_3+\cos\theta)\frac{\bar{n}_3(z_s^2+1) -(z_s^2-1)(\bar{n}_3\cos\theta+\sin^2\theta)}{4\bar{n}_3^{1/2}z_s},\\
    r_2&=\sqrt{\frac{k_0}{\e_\perp-n_\perp^2\cos^2\vf}}e^{-iS(-\vf)} \frac{(z_s^2+1)(\bar{n}_3^2+\e_\perp n_3^{(2)}\cos\theta)-\e_\perp(z_s^2-1)(n_3^{(2)}+\cos\theta)}{4\e_\perp^{1/2}(n_3^{(2)})^{1/2} z_s},\\
    l_1&=is\sqrt{\frac{k_0}{\e_\perp-n_\perp^2\cos^2\vf}}(\bar{n}_3-\cos\theta)\frac{\bar{n}_3(z_s^2+1) -(z_s^2-1)(\bar{n}_3\cos\theta-\sin^2\theta)}{4\bar{n}_3^{1/2}z_s},\\
    l_2&=-\sqrt{\frac{k_0}{\e_\perp-n_\perp^2\cos^2\vf}}e^{iS(-\vf)} \frac{(z_s^2+1)(\bar{n}_3^2-\e_\perp n_3^{(2)}\cos\theta)+\e_\perp(z_s^2-1)(n_3^{(2)}-\cos\theta)}{4\e_\perp^{1/2}(n_3^{(2)})^{1/2} z_s},
\end{split}
\end{equation}
where $\bar{n}_3:=\bar{k}_3/k_0$ and $n^{(2)}_3:=k^{(2)}_3/k_0$. The exact expressions for the coefficients $\nu_{k}$, $k=\overline{0,4}$, entering the normalization constant \eqref{norm_const} are
\begin{fleqn}
\begin{equation}
\begin{split}
    \nu_0=\,&\big[8\bar{n}^2_3(n_3^{(2)})^2(1-n_\perp^2)\e_\perp^2(4\e_\perp z_s^2-n_\perp^2(z_s^2+1)^2)^2 \big]^{-1} \bigg\{\e_\perp^4(n_3^{(2)})^4\bar{n}_3^2[4z^2_s-n_\perp^2(z_s^2+1)^2]^2+\\
    &+\bar{n}_3^2\big[\e_\perp^2(4z^2_s+n_\perp^2(z^2_s-1)^2) -2\e_\perp n_\perp^2(z^2_s+1)+n_\perp^4(z_s^4+1)^2\big]^2+\\ &+\e_\perp^2(n_3^{(2)})^2\Big[ \e_\perp^4\big(4z_s^2+n_\perp^2(z_s^2-1)^2\big)^2 -8\e_\perp^3\big[n_\perp^4(z_s^4-1)^2+2n_\perp^2(z_s^6+14z_s^4+z_s^2) -24z_s^4\big]+\\
    &+\e_\perp^2\big[6n_\perp^6(z_s^4-1)^2+n_\perp^4(13z_s^8+124z_s^6+414z_s^4+124z_s^2+13) -8n_\perp^2z_s^2(13z_s^4+50z_s^2+13)+16z_s^4\big]-\\
    &-2\e_\perp n_\perp^2\big[n_\perp^4(z_s^2+1)^4(13z_s^4+58z_s^2+13) -4n_\perp^2(z_s^2+1)^2(z_s^4+18z_s^2+1) -16z_s^4 \big]+\\ &+14n_\perp^8(z_s^2+1)^4 -8n_\perp^6(z_s^2+1)^2(z_s^4+5z_s^2+1) +16n_\perp^4z_s^4 \Big] \bigg\},
\end{split}
\end{equation}
\begin{equation}
\begin{split}
    \nu_1=\,&\frac{\chi_\perp n_\perp^4(z_s^4-1)(\bar{n}_3+\e_\perp n_3^{(2)})}{8\bar{n}_3n^{(2)}_3(1-n_\perp^2)\e_\perp(4z_s^2\e_\perp -n_\perp^2(z_s^2+1)^2)^2}\big[\e_\perp^2(1-n_\perp^2)(z_s^2-1)^2 -\bar{n}_3^4(z_s^2+1)^2+\\
    &+\chi_\perp n_\perp^2(z_s^4-1)(\bar{n}_3+\e_\perp n_3^{(2)}) +\bar{n}_3n_3^{(2)}\e_\perp(4z_s^2-n_\perp^2(z_s^2+1)^2) \big],
\end{split}
\end{equation}
\begin{equation}
    \nu_2=-\frac{\chi_\perp^2 \big[4z_s^2\bar{n}_3^2+\e_\perp n_\perp^2(z_s^2-1)^2-2\bar{n}_3n_\perp^2(z_s^4-1) \big]\big[4z_s^2\bar{n}_3^2+\e_\perp n_\perp^2(z_s^2-1)^2 \big]}{16\bar{n}_3^2(1-n_\perp^2)(4z_s^2\e_\perp -n_\perp^2(z_s^2+1)^2)^2},
\end{equation}
\begin{equation}
\begin{split}
    \nu_3=\,&-\frac{\big[\bar{n}_3^4(z_s^4+1)^2-\e_\perp^2(n_3^{(2)})^2(4z_s^2-n_\perp^2(z_s^2+1)^2)-\e_\perp^2(1-n_\perp^2)(z_s^2-1)^2 \big]}{16(n^{(2)}_3)^2(1-n_\perp^2)\e_\perp^2(4z_s^2\e_\perp -n_\perp^2(z_s^2+1)^2)^2}\times\\
    &\times \big[\bar{n}_3^4(z_s^4+1)^2-\e_\perp^2(n_3^{(2)})^2(4z_s^2-n_\perp^2(z_s^2+1)^2)-\e_\perp^2(1-n_\perp^2)(z_s^2-1)^2 -2n_3^{(2)}n_\perp^2\chi_\perp\e_\perp(z_s^4-1) \big],
\end{split}
\end{equation}
\begin{equation}
\begin{split}
    \nu_4=\,&\frac{\chi_\perp n_\perp^2(\bar{n}_3-\e_\perp n_3^{(2)})(z_s^4-1)}{8\bar{n}_3n_3^{(2)}(1-n_\perp^2)\e_\perp(4z_s^2\e_\perp-n_\perp^2(z_s^2+1)^2)^2}
    \big[\bar{n}_3 (z_s^2+1)^2(\e_\perp n_\perp^2 n_3^{(2)}-\bar{n}_3^3)-\\
    &-\chi_\perp n_\perp^2(z_s^4-1) (\bar{n}_3-\e_\perp n_3^{(2)}) -4\bar{n}_3n_3^{(2)}\e_\perp z_s^2 +(1-n_\perp^2)(z_s^2-1)^2\e_\perp^2\big],
\end{split}
\end{equation}
\end{fleqn}
where, for brevity, we denote $z_s:=e^{is\vf}$. As is seen, the dependence of $\nu_k$ on $s$ and $\vf$ is gathered into the combination $s\vf$.

\section{Paraxial approximation}\label{C}

In the leading order in $n_\perp^2$, the solutions \eqref{ordinary_+}-\eqref{extraordinary_+} are written as follows: the direct ordinary wave
\begin{equation}\label{semiclass_paraxa1}
    a^{(1)}_\pm=\pm i(2\bar{k}_0)^{-1/2} e^{i\bar{k}_3z \pm i\bar{\theta}},\qquad
    A_3^{(1)}=\frac{k_\perp\sin\bar{\theta}}{(2\bar{k}_0^3)^{1/2}}e^{i\bar{k}_3z};
\end{equation}
the reflected ordinary wave
\begin{equation}
    \tilde{a}^{(1)}_\pm=\pm i(2\bar{k}_0)^{-1/2} e^{-i\bar{k}_3z \pm i\bar{\theta}},\qquad
    \tilde{A}_3^{(1)}=-\frac{k_\perp\sin\bar{\theta}}{(2\bar{k}_0^3)^{1/2}}e^{-i\bar{k}_3z};
\end{equation}
the direct extraordinary wave
\begin{equation}
    a^{(2)}_\pm=
    \big(2k_0\e_\parallel^{1/2}\big)^{-1/2}e^{iS(\bar{\theta}) \pm i\bar{\theta}},\qquad A_3^{(2)}=-k_\perp\cos\bar{\theta}\bigg(\frac{k_0\e_\parallel^{1/2}}{2\bar{k}_0^4}\bigg)^{1/2} e^{iS(\bar{\theta})};
\end{equation}
the reflected extraordinary wave
\begin{equation}\label{semiclass_paraxb2}
    \tilde{a}^{(2)}_\pm=
    \big(2k_0\e_\parallel^{1/2}\big)^{-1/2}e^{-iS(\bar{\theta}) \pm i\bar{\theta}},\qquad \tilde{A}_3^{(2)}=k_\perp\cos\bar{\theta} \bigg(\frac{k_0\e_\parallel^{1/2}}{2\bar{k}_0^4}\bigg)^{1/2} e^{-iS(\bar{\theta})}.
\end{equation}
Notice that we do not develop the arguments of the fast oscillating exponents as a series in $n_\perp$ and neglect the small terms only in the preexponential factors. In order to discard the contributions proportional to $n^2_\perp$ in the fast oscillating exponents, the more stringent conditions for large $L$ must be satisfied,
\begin{equation}
    \e_\perp^{-1/2}k_0n_\perp^2 L\ll1,\qquad \sqrt{1+\de\e}\e_\perp^{-1/2}k_0n_\perp^2 L\ll1,
\end{equation}
and not just $n_\perp^2\ll1$.

The exact expressions for $r_{1,2}$, $l_{1,2}$ are presented in \eqref{ach_wkb}. In the paraxial approximation, neglecting the contributions of the order of $n_\perp^2$ and higher, we obtain
\begin{equation}\label{ri_li}
\begin{aligned}
    r_1&=is\sqrt{\frac{k_0}{\e_\perp^{1/2}}} \frac{\e_\perp^{1/2}+1}{2z_s},&\qquad r_2&=\sqrt{\frac{k_0}{\e_\parallel^{1/2}}} \frac{\e_\parallel^{1/2}+1}{2z_s}e^{-iS(-\vf)},\\
    l_1&=is\sqrt{\frac{k_0}{\e_\perp^{1/2}}} \frac{\e_\perp^{1/2}-1}{2z_s},&\qquad l_2&=\sqrt{\frac{k_0}{\e_\parallel^{1/2}}} \frac{\e_\parallel^{1/2}-1}{2z_s}e^{iS(-\vf)}.
\end{aligned}
\end{equation}
We see that the coefficients at the reflected waves are proportional to $\chi_{\perp,\parallel}$ and, as a rule, are small. In the paraxial regime, the coefficient specifying the normalization constant \eqref{norm_const} become
\begin{equation}\label{nuk_wkb}
\begin{split}
    \nu_0=\,&\frac{1}{16}\big(12+\e_\perp+\e_\perp^{-1}+\e_\parallel+\e_\parallel^{-1} \big) -\frac{n_\perp^2\de\e}{32\e_\parallel^2\e_\perp z_s^2}\big[(\e_\parallel^2\e_\perp^2-\e_\parallel -\chi^2_\parallel\e_\perp/2)(z_s^4+1) -(\e_\parallel^2-1)\e_\perp z_s^2 \big],\\
    \nu_1=\,&n_\perp^2 \chi_\perp (z_s^4-1)  \frac{(\e_\parallel^{1/2}-\e_\perp^{1/2})(\e_\parallel^{1/2}\e_\perp^{1/2}+1)}{64z_s^2\e_\parallel^{1/2}\e_\perp^{3/2}},\\
    \nu_2=\,&-\frac{\chi_\perp^2}{32\e_\perp} -\frac{n_\perp^2\chi_\perp^2}{64\e_\perp^2 z_s^2}\big[(\e_\perp-\e_\perp^{1/2}+1)z_s^4 +\e_\perp+\e_\perp^{1/2}+1\big],\\
    \nu_3=\,&-\frac{\chi_\parallel^2}{32\e_\parallel} +\frac{n_\perp^2\chi_\parallel}{128\e_\parallel^2\e_\perp z_s^2}\Big\{(\e_\parallel^{1/2}-1)\big[(\e_\parallel^{1/2}+1)(\e_\parallel+\e_\perp) +2\e^{3/2}_\parallel\chi_\perp\big]z_s^4 -2(\e_\parallel+1)(\e_\parallel-\e_\perp)z_s^2+\\
    &+(\e_\parallel^{1/2}+1)\big[(\e_\parallel^{1/2}-1) (\e_\parallel+\e_\perp) +2\e^{3/2}_\parallel\chi_\perp\big]\Big\},\\
    \nu_4=\,& n_\perp^2 \chi_\perp (z_s^4-1)  \frac{(\e_\parallel^{1/2}+\e_\perp^{1/2})(\e_\parallel^{1/2}\e_\perp^{1/2}-1)}{64z_s^2\e_\parallel^{1/2}\e_\perp^{3/2}},
\end{split}
\end{equation}
Notice that the second term in $\nu_0$ and the expressions for $\nu_{k}$, $k=\overline{1,4}$, are of order $\chi^2_{\perp,\parallel}$, i.e., they are small. The terms proportional to $n_\perp^2$ are suppressed even stronger.


\end{document}